\documentclass[aps,nofootinbib,amsmath,prd,onecolumn,notitlepage,showpacs,superscriptaddress,groupedaddress]{revtex4-1}

\usepackage[american]{babel}
\usepackage{amsfonts}
\usepackage{amsmath}
\usepackage{bbold}
\usepackage{amssymb}
\usepackage{epsfig}
\usepackage{latexsym}
\usepackage{paralist}
\usepackage{fancyhdr}
\usepackage{graphicx}
\usepackage{young}
\usepackage{etex}
\usepackage{braket}
\usepackage{float}
\usepackage{slashed}
\usepackage{xcolor}
\usepackage{soul}
\usepackage{dcolumn} 
\usepackage{physics}
\usepackage{bm}
\usepackage{nicefrac}

\makeatletter\AtBeginDocument{\let\@elt\relax}\makeatother

\def\bea{\begin{eqnarray}} 
\def\eea{\end{eqnarray}}
\def\be{\begin{equation}} 
\def\ee{\end{equation}} 
\def\ba{\begin{array}}
\def\ea{\end{array}}

\def\be{\begin{equation}}
\def\ee{\end{equation}}
\def\bea{\begin{eqnarray}}
\def\eea{\end{eqnarray}}

\usepackage{amsmath}

\begin{document}

\title{Conformally covariant operators of mixed-symmetry tensors and MAGs}

\author{Gregorio Paci}
\email{gregorio.paci@phd.unipi.it}
\affiliation{
Universit\`a di Pisa, Largo Bruno Pontecorvo 3, 56127 Pisa, Italy}
\affiliation{INFN - Sezione di Pisa, Largo Bruno Pontecorvo 3, 56127 Pisa, Italy}

\author{Dario Sauro}
\email{dario.sauro@phd.unipi.it}
\affiliation{
Universit\`a di Pisa, Largo Bruno Pontecorvo 3, 56127 Pisa, Italy}
\affiliation{INFN - Sezione di Pisa, Largo Bruno Pontecorvo 3, 56127 Pisa, Italy}

\author{Omar Zanusso}
\email{omar.zanusso@unipi.it}
\affiliation{
Universit\`a di Pisa, Largo Bruno Pontecorvo 3, 56127 Pisa, Italy}
\affiliation{INFN - Sezione di Pisa, Largo Bruno Pontecorvo 3, 56127 Pisa, Italy}

\begin{abstract}
%
We compute conformally covariant actions and operators for tensors with mixed symmetries in arbitrary dimension $d$. Our results complete the classification of conformal actions that are quadratic on arbitrary tensors with three indices, which allows to write corresponding conformal actions for all tensor species that appear in the decomposition of the distorsion tensor of an arbitrary metric-affine theory of gravity including both torsion and nonmetricity.
We also discuss the degrees of freedom that such theories are propagating, as well as interacting metric-affine theories that enjoy the conformal actions in the Gaussian limit.
\end{abstract}

\maketitle

\renewcommand{\thefootnote}{\arabic{footnote}}
\setcounter{footnote}{0}

\section{Introduction} \label{sect:intro}

Tensors with mixed-symmetries arise as irreducible tensorial representations of the group $GL(d)$ whenever we have more than two indices. These tensors have attracted an increasing amount of attention over the past years, despite being known to be interesting and useful for quite some time \cite{Curtright:1980yk}.

From the point of view of the geometry of General Relativity and its metric-affine extensions \cite{Kibble:1961ba,Charap:1973fi,Hehl:1994ue,Gronwald:1995em,BeltranJimenez:2018vdo,BeltranJimenez:2019esp,Percacci:2023rbo}, the simplest mixed-symmetry tensors emerge as irreducible components in the tensor decomposition of torsion and nonmetricity, that is, for independent connections that bring more degrees of freedom than the Levi-Civita one. The reason is that a general $GL(d)$ connection depends on a three-indexed tensor, which is enough to contain the first two nontrivial mixed-symmetric irreducible representations,
known as the traceless hook-symmetric and hook-antisymmetric tensors.
From the point of view of conformal field theory, mixed-symmetry tensors of arbitrary rank appear in the block-decomposition of four-point functions
built with operators with various spins \cite{Costa:2014rya,Costa:2016hju}.
In both contexts, the analysis of mixed-symmetry tensors represent a rather
cumbersome departure from the simpler examples of tensors
that appear up to spin-two, including scalars, vectors and traceless-symmetric ones.

In this paper we try to connect the two aforementioned ``worlds'' -- metric-affine extensions of gravity and conformal field theory -- by discussing how the simplest mixed-symmetry tensors appear in metric-affine theories and what are their conformal properties. Even tough scale and conformal invariance have already been studied in the metric-affine context (see, e.g. \cite{Karananas:2015ioa,Iosifidis:2018zwo,Karananas:2021gco}), previous analyses have not dealt neither with the conformal coupling of mixed-symmetry tensors, nor with terms that are expected to be relevant in the ultraviolet (UV) regime. To this end, we construct curved-space conformally covariant actions and operators for the hook-antisymmetric and hook-symmetric traceless tensors that appear as components in the tensorial decomposition of a general distorsion of
the Levi-Civita connection. In the flat space limit, the conformally covariant actions become models of free conformal field theories for tensors with mixed-symmetries.

We then discuss some properties of the newly found actions and operators, paying particular attention to the spin degrees of freedom that they propagate. We also discuss more general metric-affine theories of gravity that can be interpreted as self-interacting versions of the conformal actions. Our hope is to draw a meaningful connection between two rather different areas of research, i.e.\ metric-affine theories and conformal field theory, since this connection could prove useful in providing ultraviolet completions of General Relativity \cite{Pagani:2015ema,Gies:2022ikv} (as they could be expected to be conformally invariant) and interesting new degrees of freedom for future phenomenological applications.

The hasty reader can find the highlights of the tensor decomposition of a general metric-affine theory of gravity in Sect.~\ref{sect:introduction},
and the derivation of the conformal actions
in Sects.~\ref{sect:hook-antisymmetric} and \ref{sect:hook-symmetric}.
The relations of the conformal actions with interacting metric-affine theories
are discussed in Sect.~\ref{sect:MaptoMags} and \ref{sect:MaptoMags-2}.
Throughout the paper we take some detours to highlight interesting aspects and make contact with the literature.

\section{A crash course on the degrees of freedom of MAGs} \label{sect:introduction}

In metric-affine theories of gravity (MAGs) the metric $g_{\mu\nu}$ and the connection $\nabla$ are regarded as independent variables on a $d$-dimensional manifold. The connection $\nabla$ is a $GL(d)$ connection, in $d$ spacetime dimensions, that can, in general, be nonsymmetric and noncompatible with the metric, as opposed to the Levi-Civita connection $\mathring{\nabla}$ of General Relativity (GR) which is the unique symmetric and compatible connection.\footnote{%
Quantities based on the Levi-Civita connection are often denoted with the small ring in the MAG literature to distinguish them from the same quantities obtained from the general connection. We follow this convention, even if later on we are mostly working only with the Levi-Civita connection.
} 
Even though the original formulation of MAGs employed non-holonomic indices in order to highlight the analogies with non-abelian gauge theories (see, e.g., \cite{Hehl:1976kj,Hehl:1976my}), here we choose to work with coordinate indices. Such a choice does not jeopardize our results in any way, since it only amounts to a gauge choice \cite{Percacci:2020bzf}. Moreover, calculations are less demanding in this framework and connections to the literature of conformal field theories are easier to make.
While the geometrical content of a MAG is potentially much more complicate than that of a metric's theory, it is always possible to split the components of the independent connection
\begin{equation}\label{eq:distorsion-def}
 \begin{split}
  \Gamma^\mu{}_{\nu\rho} = \mathring{\Gamma}^\mu{}_{\nu\rho} + \Phi^\mu{}_{\nu\rho}\,,
 \end{split}
\end{equation}
where $\Phi$ is a tensor with three indices, being the difference of two connections, and is known as distorsion. The nature of the distorsion tensor can be further quantified by introducing torsion and nonmetricity
\begin{equation}\label{eq:torsion-nonmetricity-def}
 \begin{split}
  T^\mu{}_{\nu\rho} = 2\Gamma^\mu{}_{[\rho\nu]} \,, \qquad Q_{\mu\nu\rho}=-\nabla_\mu g_{\nu\rho}\,,
 \end{split}
\end{equation}
which can be naively seen as the curvatures of the frame and of the metric fields, respectively.
It is then relatively easy to show that $\Phi$ is the sum of two terms
\begin{equation}\label{eq:distortion-decomposition}
 \begin{split}
  \Phi^\mu{}_{\nu\rho} = K^\mu{}_{\nu\rho} + L^\mu{}_{\nu\rho}\,,
 \end{split}
\end{equation}
known as contortion and disformation. The contortion is a function only of the torsion
\begin{equation}\label{eq:contortion-def}
 \begin{split}
  K^\mu{}_{\nu\rho} = \frac{1}{2}\Bigl(T_\rho{}^\mu{}_\nu+T_\nu{}^\mu{}_\rho-T^\mu{}_{\nu\rho}\Bigr) \,,
 \end{split}
\end{equation}
while the disformation is a function only of the nonmetricity
\begin{equation}\label{eq:disformation-def}
 \begin{split}
  L^\mu{}_{\nu\rho} = \frac{1}{2}\Bigl(Q_\rho{}^\mu{}_\nu+Q_\nu{}^\mu{}_\rho-Q^\mu{}_{\nu\rho}\Bigr) \,.
 \end{split}
\end{equation}

It is straightforward to recover the Levi-Civita limit adopted in General Relativity by requiring $T=Q=0$ (compatible and symmetric connection), which implies $\Phi=0$ and reduces the affine structure to $\nabla=\mathring{\nabla}$. Since the Levi-Civita connection is completely determined from the derivatives of the metric, $2\mathring{\Gamma}^\mu{}_{\nu\rho}=g^{\mu\theta}(\partial_{\nu} g_{\rho\theta}+\partial_{\rho} g_{\theta\nu}-\partial_{\theta} g_{\rho\nu})$, MAGs can sometimes be understood as linearized theories if compared to metric models, though their field content is generally much more complicate,
both in terms of interactions and of fields with spin \cite{Baldazzi:2021kaf}.
Interestingly, however, both the Cartan's and Palatini's formulations of General Relativity with the Einstein-Hilbert action, which are restricted to respectively contain only torsion or nonmetricity, can be shown to be dynamically equivalent to the metric formulation when the equations of motion of the connection are used \cite{Dadhich:2012htv,Trautman:2006fp}.

Obviously, the connection $\nabla$ has its own curvature in general, and we denote it as $R^\mu{}_{\nu\rho\theta} v^\nu = [\nabla_\rho,\nabla_\theta] v^\mu + T^\lambda{}_{\rho\theta} \nabla_\lambda v^\mu$,
even though, strictly speaking, it is not Riemannian. Similarly, $\mathring{R}^\mu{}_{\nu\rho\theta} v^\nu = [\mathring{\nabla}_\rho,\mathring{\nabla}_\theta] v^\mu$ is the standard Riemann curvature in our notation, and the two curvatures are related
\begin{equation}
 \begin{split}
  R^\mu{}_{\nu\rho\theta} = \mathring{R}^\mu{}_{\nu\rho\theta}
  + \mathring{\nabla}_\rho\Phi^\mu{}_{\nu\theta}
  -\mathring{\nabla}_\theta\Phi^\mu{}_{\nu\rho} + \Phi^\mu{}_{\eta\rho} \Phi^\eta{}_{\nu\theta}
  -\Phi^\mu{}_{\eta\theta} \Phi^\eta{}_{\nu\rho}\,.
 \end{split}
\end{equation}
It is thus possible to write any MAG action as a function of $(\nabla,T,Q)$
or of $(\mathring{\nabla},T,Q)$, which we nickname the Cartan's and the Einstein's forms respectively \cite{Baldazzi:2021kaf}.

\subsection{General tensor decompositions of torsion and nonmetricity} \label{sect:tensors}

The disformation tensor, as well as all the other tensors defined in the previous section, introduce several new degrees of freedom in gravitational actions of geometrical origin. Using a spin decomposition, one can see that both torsion and nonmetricity contain spin-$2$ degrees of freedom with the correct parity, so it is possible to write geometrical alternatives to Einstein's theory that are dynamically equivalent to GR, but based entirely on torsion and nonmetricity, known respectively as antisymmetric and symmetric teleparallel gravities \cite{BeltranJimenez:2018vdo,BeltranJimenez:2019esp}.

In this paper we are more concerned with tensor decompositions, rather than spin decompositions. In order to decompose these tensors, we start by defining their nontrivial traces as
\begin{equation}
 \tau_\mu \equiv T^\rho{}_{\mu\rho} \, , \qquad Z_\mu \equiv Q_\mu{}^\rho{}_\rho \, , \qquad V_\mu \equiv Q_{\rho\mu}{}^\rho \, ,
\end{equation}
where the first trace is commonly known as torsion-vector and a combination of $Z_\mu$ and $V_\mu$ is usually referred to as the Weyl vector. The Weyl vector has to do with a gauged version of Weyl symmetry, which is interesting, but it is not what we go about in this paper, so we refer to Appendix~\ref{sect:relation} for the fleshing out of the relations with that approach. A further step is provided by singling out the completely antisymmetric and completely symmetric tracefree components
\begin{equation}
 H_{\mu\nu\rho} \equiv \frac{1}{3} T_{[\mu\nu\rho]} \, , \qquad C_{\mu\nu\rho} \equiv \frac{1}{3} \Bigl\{ Q_{(\mu\nu\rho)} - \frac{1}{d+2} \Bigl( B_\mu g_{\nu\rho} + B_\nu g_{\rho\mu} + B_\rho g_{\mu\nu} \Bigr)  \Bigr\}\, ,
\end{equation}
where we have introduced a shorthand for the combination $B_\mu = 2 Z_\mu + V_\mu$.
Therefore, the torsion and nonmetricity tensors $T^\mu{}_{\nu\rho}$ and $Q_{\mu\nu\rho}$ can be decomposed as
\begin{eqnarray}\label{eq:T-Q-decomposition}
  T^\mu{}_{\nu\rho} &=& \frac{1}{d-1} \left(\delta^\mu{}_\rho \tau_\nu - \delta^\mu{}_\nu \tau_\rho \right)
  + H^\mu{}_{\nu\rho}
  +\kappa^\mu{}_{\nu\rho} \, ,
  \\
  Q_{\mu\nu\rho} &=&
  \frac{d}{d^2+d-2}\Bigl(g_{\mu\rho} V_\nu+g_{\mu\nu} V_\rho-\frac{2}{d}g_{\nu\rho}V_\mu \Bigr)
  +\frac{d+1}{d^2+d-2} \Bigl(g_{\nu\rho} Z_\mu - \frac{1}{d+1}g_{\mu\nu} Z_\rho-\frac{1}{d+1} g_{\mu\rho} Z_\nu  \Bigr)
  + C_{\mu\nu\rho} + \psi_{\mu\nu\rho}\,. \nonumber
\end{eqnarray}
For the decomposition of the torsion tensor and extended discussions on its relation to conformal transformations, we refer to \cite{Shapiro:2001pat} and references therein. We are not aware of similar discussions involving the decomposition of nonmetricity, besides its relation with Weylian geometry, which we address in appendix~\ref{sect:relation}.
In short, we have the three vectors that appear as the nontrivial traces of $T$ and $Q$, but also $H_{\mu\nu\rho}$, which is a totally antisymmetric three-tensor (i.e., a three-form), and $C_{\mu\nu\rho}$, which is a totally symmetric traceless tensor ($C^\mu{}_{\mu\nu}=0$). Most importantly for us, the tensor decompositions include two mixed-symmetry tensors:
a hook-antisymmetric traceless tensor $\kappa^\mu{}_{\nu\rho}$ that satisfies
\begin{equation}
 \begin{split}
  \kappa^\mu{}_{\mu\rho}=0\,, \qquad \kappa^\mu{}_{[\nu\rho]}=\kappa^\mu{}_{\nu\rho}\,, \qquad \kappa_{[\mu\nu\rho]}=\frac{1}{3}\left( \kappa_{\mu\nu\rho}+\kappa_{\nu\rho\mu}+\kappa_{\rho\mu\nu}\right)=0 \,,
 \end{split}
\label{HAproperties}
\end{equation}
and a hook-symmetric traceless tensor $\psi^\mu{}_{\nu\rho}$ that satisfies
\begin{equation}
 \begin{split}
  \psi^\mu{}_{\mu\rho}=0\,, \qquad g^{\nu\rho} \psi^\mu{}_{\nu\rho}=0\,, \qquad \psi^\mu{}_{(\nu\rho)}=\psi^\mu{}_{\nu\rho}\,, \qquad  \psi_{(\mu\nu\rho)}=\frac{1}{3}\left(\psi_{\mu\nu\rho}+\psi_{\nu\rho\mu}+\psi_{\rho\mu\nu}\right)=0 \,.
 \end{split}
\label{HSproperties}
\end{equation}
In four dimensions all these tensors are present and largely unconstrained, with the exception that the totally antisymmetric tensor $H$ is generally replaced by its dual axial vector. In general dimension $d>2$ the spaces of hook-symmetric and hook-antisymmetric tensors have the same dimensionality, that is, $\frac{1}{3}d(d+2)(d-2)$.

The suggestion, pushed forward in Refs.~\cite{Sauro:2022chz,Sauro:2022hoh},
is to constrain the degrees of freedom of a general MAG using symmetry requirements, and in particular Weyl and conformal invariance. In particular,
while the vectors in the expansions of both torsion and nonmetricity can be used to construct models with gauged Weyl symmetry, the hook-symmetric and antisymmetric tensors can enjoy fully conformal invariant actions without the need of gauging the covariant derivative, provided that $\kappa$ and $\psi$ are dimensionless fields.

Of course, in order to have a complete understanding of the conformal behavior of metric-affine theories one should also study the conformal coupling of matter fields to metric-affine geometries, as it has been done, e.g., in \cite{Iosifidis:2018zwo}. On the other hand, our present analysis is clearly valid regardless of such an issue. Indeed, before addressing the problem of constructing a conformal theory including matter fields, we still have to built up one in the gravitational sector. For the sake of completeness, however, let us make some remarks about the coupling of the theories we are examining with matter fields. Mixed symmetry tensors can be coupled in a conformal way to spin-$0$ matter only through potential terms that are quadratic both in the tensor and scalar field, while they cannot be coupled in such a way to spin-$\frac{1}{2}$ fields. In passing, we note that an interaction term like $\mathcal{L} \supset \overline{\Psi}_\mu \gamma_{[\nu} \Psi_{\rho]} \kappa^{\rho\mu\nu}$ between a Rarita-Schwinger spin-$\frac{3}{2}$ fermion \cite{rarita1941theory} and $\kappa$ is conceivable and it is Weyl invariant \emph{only} in $d=4$. Obviously, this is just one particular example of interaction terms between Rarita-Schwinger fermions and mixed-symmetry tensors, and the coupling constants of such interactions are free parameters of the theory. Finally, we have to say that our subsequent analysis is not definitive as far as the most general MAG is concerned. This is because, as we have just seen, in MAGs there are many tensor fields to account for, and it is non-trivial to obtain the complete and interacting conformal coupling of all of them.

In the next section we discuss conformal actions based on the fields of the decompositions and derive new actions for $\kappa$ and $\psi$.

\section{The conformal actions} \label{sect:conformal-actions}

It should be clear from the content of the decompositions of the previous section
that a completely general MAG theory allows for many possible interaction terms. There can be almost a thousand terms if all interactions with up to four derivatives are considered (the equivalent of $R^2$ gravity) \cite{Baldazzi:2021kaf}. Therefore, a guiding principle based on symmetry could greatly help in constraining the number of possible interactions. In this respect, conformal symmetry is generally the most constraining symmetry that could be asked for. In curved space we naturally refer to symmetry under the Weyl rescalings of the metric by a local conformal factor, $g_{\mu\nu} \to g'_{\mu\nu}=\Omega^2 g_{\mu\nu}$, which is known to imply invariance under the conformal group that extends the Poincar\'e group in flat spacetime.

The conformal transformation of the field species involved in the decompositions
can be inferred on the basis of scaling considerations and thus they are tied to the number of derivatives involved in the actions. The fields in \eqref{eq:T-Q-decomposition} are naturally attributed a Weyl weight $(4-d)/2$ for a conformal action involving two derivatives, therefore for consistency with the MAG interpretation of the degrees of freedom the limit $d=4$ is implied. Nevertheless, we construct general $d$-dependent conformal actions as they can be useful in other contexts and teach us important properties of the conformal theories, as will become clear below.

Conformal actions for $p$-forms and general traceless-symmetric rank-$p$ tensors
have already been constructed in the literature in complete generality \cite{Erdmenger:1997gy,Erdmenger:1997wy,Osborn:2016bev}. These thus cover all the tensors involved in \eqref{eq:T-Q-decomposition} with the notable exceptions of the mixed-symmetry $\kappa$ and $\psi$. 
As far as we know, the construction of conformal actions and operators on tensors with hook symmetry such as $\kappa^\mu{}_{\nu\rho}$ and $\psi^\mu{}_{\nu\rho}$ has not been carried out in full generality before. Indeed, although many works have discussed conformal properties of these types of tensors, these properties have been restricted to particular backgrounds. In particular, they were studied on Bach flat \cite{Ponds:2005bfb} and conformally flat spaces \cite{Ponds:1902cghs} for both $\kappa^\mu{}_{\nu\rho}$ and $\psi^\mu{}_{\nu\rho}$, on conformally flat Einstein spaces \cite{Queva:2015cghs} for $\psi^\mu{}_{\nu\rho}$. Actually, a complete classification of conformal operators was given on Minkowski spacetime in Refs.~\cite{Vasiliev:2009bchs, Vasiliev:2006cde, Grigoriev:2015cin}. Further interesting work on the subject is Ref.~\cite{Grigoriev:2020cffg}, where the authors use the ambient space formalism, which, however, is restricted to the Minkowski metric on the boundary, although it is clear that such approach could be extended to conformally flat boundary metrics too. In this paper, we construct conformal actions and operators for $\kappa^\mu{}_{\nu\rho}$ and $\psi^\mu{}_{\nu\rho}$ without any assumption on the background. To do so, we tackle this problem following a rather general procedure introduced by Erdmenger and Osborn \cite{Erdmenger:1997wy} in Sects.~\ref{sect:hook-antisymmetric} and \ref{sect:hook-symmetric}.

For illustrative purposes, we first briefly present in Sect.~\ref{sect:conformal-past} the conformal action of a vector field with two derivatives, which reduces to the Maxwell action in $d=4$.

\subsection{Warm up example: the conformal vector form} \label{sect:conformal-past}

We concentrate on conformal actions that are quadratic in the fields,
which thus interact only with the local curvatures expressed in the Levi-Civita basis, and become standard free theories in the flat space limit. The simplest example that is relevant for our analysis is thus the conformal action of a vector form, denoted $A_\mu$, which can be written as
\begin{equation}
 \begin{split}
  S_{\rm c}[g,A] &= \frac{1}{2} \int {\rm d}^dx \sqrt{g} \Bigl\{
   \mathring{\nabla}_\mu A_\nu \mathring{\nabla}^\mu A^\nu
  -\frac{4}{d} (\mathring{\nabla}_\mu A^\mu)^2
  +2 \mathring{K}_{\mu\nu} A^\mu A^\nu + \frac{d-2}{2} \mathring{{\cal J}} A^2
  \Bigr\}\,.
 \end{split}
\end{equation}
We have that $S_{\rm c}[g,A]$ is invariant under the conformal transformation
$g_{\mu\nu} \to g'_{\mu\nu}=\Omega^2 g_{\mu\nu}$ and $A_\mu \to A'_\mu=\Omega^{\frac{4-d}{2}}A_\mu$.
In $S_{\rm c}[g,A]$ we introduced the Schouten tensor and its trace
\begin{equation}\label{eq:def-schouten}
 \begin{split}
  \mathring{K}_{\mu\nu} = \frac{1}{d-2}\Bigl(\mathring{R}_{\mu\nu}-\frac{1}{2(d-1)}g_{\mu\nu} \mathring{R}\Bigr)\,, \qquad \mathring{{\cal J}}=\frac{1}{2(d-1)}\mathring{R}\,.
 \end{split}
\end{equation}
The Schouten tensor is particularly well-suited for the analysis of conformally invariant theories because under an infinitesimal Weyl transformation $\Omega=1+\sigma$, $\delta_\sigma^W g_{\mu\nu}= 2 \sigma g_{\mu\nu}$, it transforms as
$
\delta_\sigma^W \mathring{K}_{\mu\nu} = -\mathring{\nabla}_\mu \partial_\nu \sigma
$,
so it can be used to compensate derivatives of the conformal parameter $\sigma$
coming from the transformation of derivative terms.
The poles in the definitions of $\mathring{K}$ and $\mathring{{\cal J}}$
play the role of obstructions to the lift of flat space's conformal invariance to Weyl invariance in lower dimensionalities \cite{Fefferman:2007rka}. The use of $\mathring{K}$ and $\mathring{{\cal J}}$ greatly simplifies the form of conformal actions for reasons that will become clearer also in the next subsections.
For later purpose, we also introduce the Weyl tensor, which can be elegantly expressed in terms of the Riemann and Schouten tensors as
\begin{equation}\label{eq:Riem-to-Weyl&Schouten}
 \mathring{W}_{\mu\nu\alpha\beta} = \mathring{R}_{\mu\nu\alpha\beta} - \Bigl( g_{\mu\alpha} \mathring{K}_{\nu\beta} - g_{\nu\alpha} \mathring{K}_{\mu\beta} + g_{\nu\beta} \mathring{K}_{\mu\alpha} - g_{\mu\beta} \mathring{K}_{\nu\alpha} \Bigr) \, ,
\end{equation}
and is conformally invariant if one index is raised with the metric. The Weyl tensor does not appear in $S_{\rm c}[g,A]$ simply because the vector $A_\mu$ does not have enough indices to produce a conformally invariant contraction, but it does appear in generalizations involving fully antisymmetric \cite{Erdmenger:1997gy} and traceless symmetric tensors \cite{Erdmenger:1997wy}.


One amusing property of $S_{\rm c}[g,A]$ becomes evident if we rewrite the first two derivative terms using the exterior differential $F_{\mu\nu}=\partial_\mu A_\nu-\partial_\nu A_\mu$. One finds that, up to a total derivative,
\begin{equation}
 \begin{split}
  S_{\rm c}[g,A] &= \int {\rm d}^dx \sqrt{g} \Bigl\{
   \frac{1}{4} F^{\mu\nu} F_{\mu\nu}
  +\frac{d-4}{2d} (\mathring{\nabla}_\mu A^\mu)^2
  -\frac{d-4}{2} \mathring{K}_{\mu\nu} A^\mu A^\nu + \frac{d-4}{4} \mathring{{\cal J}} A^2
  \Bigr\}\,,
 \end{split}
\end{equation}
from which it becomes evident that only in $d=4$ the action $S_{\rm c}[g,A]$ enjoys both
conformal and gauge invariance under the transformation $A_\mu \to A'_\mu=A_\mu-\partial_\mu \lambda$ (i.e., it becomes the Maxwell action \cite{El-Showk:2011xbs}).
From $S_{\rm c}[g,A]$ it is straightforward to obtain a conformally covariant rank two operator, ${\Delta}$, that acts on vectors
\begin{equation}
 \begin{split}
  (\Delta \cdot A)_\mu &= \frac{g_{\mu\nu}}{\sqrt{g}}\frac{\delta S_{\rm c}[g,A]}{\delta A_\nu}
  = -\mathring{\nabla}^2 A_\mu + \frac{4}{d}\mathring{\nabla}^\nu \mathring{\nabla}_\mu A_\nu
  -\frac{2(d-4)}{d} \mathring{K}_\mu{}^\nu A_\nu + \frac{(d+2)(d-4)}{2d}\mathring{{\cal J}} A_\mu\,,
 \end{split}
\end{equation}
which is a function of $g_{\mu\nu}$ only and transforms with a conformal bi-weight
\begin{equation}
 \begin{split}
 \Delta \to \Delta' =  \Omega^{-\frac{d+4}{2}} \Delta \Omega^{\frac{d-4}{2}}
 \end{split}
\end{equation}
under conformal transformations.
The generalizations of $S_{\rm c}[g,A]$ and of $\Delta$ for totally antisymmetric tensors and traceless symmetric tensors of any rank are also known, and enjoy similar features, besides having enough indices to couple to the Weyl tensor. It is therefore already known how to construct a quadratic conformal action for the tensors $H_{\mu\nu\rho}$ and $C_{\mu\nu\rho}$ that appeared in the previous section.
In the next two subsections we finally turn to the construction of conformal actions and operators on $\kappa^\mu{}_{\nu\rho}$ and $\psi^\mu{}_{\nu\rho}$.

\subsection{Conformal coupling of the hook-antisymmetric tensor} \label{sect:hook-antisymmetric}

The natural Weyl weight of the hook-antisymmetric $\kappa^\mu{}_{\nu\rho}$ tensor is $w_\kappa=\frac{4-d}{2}$, which can be derived with a simple rigid infinitesimal transformation. Notice that raising and lowering the indices with the metric does change the weight, so we establish that the default covariance of $\kappa^\mu{}_{\nu\rho}$ is the one of a rank-$(1,2)$ tensor.

The main actors are the field $\kappa^\mu{}_{\nu\rho}$, the Schouten tensor defined in \eqref{eq:def-schouten} and its trace, which transforms as
\begin{equation}\label{eq:Weyl-var-Rk}
 \delta^W_\sigma \kappa^\rho{}_{\mu\nu} = - \frac{d-4}{2} \sigma \kappa^\rho{}_{\mu\nu} \, , \qquad \delta^W_\sigma \mathring{K}_{\mu\nu} = -\mathring{\nabla}_\nu \partial_\mu \sigma\, , \qquad \delta^W_\sigma \mathring{\mathcal{J}} =-2\sigma \mathring{\mathcal{J}}  -\mathring{\square} \sigma\, .
\end{equation}
under an infinitesimal Weyl transformation of the metric.
Using the transformation of $\kappa^\mu{}_{\nu\rho}$ and of the Levi-Civita connection $\delta^W_\sigma \mathring{\Gamma}^\rho{}_{\mu\nu}=\left( \delta^\rho{}_\mu \partial_\nu \sigma + \delta^\rho{}_\nu \partial_\nu \sigma - g_{\mu\nu} \partial^\rho \sigma \right)$, we find the transformation of the covariant derivatives
\begin{equation}
\begin{split}
 \delta^W_\sigma \mathring{\nabla}_\lambda \kappa^\rho{}_{\mu\nu} = & - \frac{d-2}{2} \kappa^\rho{}_{\mu\nu} \partial_\lambda \sigma + \delta^\rho{}_\lambda \kappa^\alpha{}_{\mu\nu} \partial_\alpha \sigma - \kappa_{\lambda\mu\nu} \partial^\rho \sigma - \kappa^\rho{}_{\lambda\nu} \partial_\mu \sigma\\
 & \, + g_{\mu\lambda} \kappa^{\rho\alpha}{}_\nu \partial_\alpha \sigma - \kappa^\rho{}_{\mu\lambda} \partial_\nu \sigma + g_{\nu\lambda} \kappa^\rho{}_\mu{}^\alpha \partial_\alpha \sigma \,.
 \end{split}
\end{equation}
A rather lengthy analysis that uses the hook-antisymmetric symmetries given in Eq.~\eqref{HAproperties} reveals that there are five independent contractions of the form $(\mathring{\nabla}\kappa)^2$ (see Appendix \ref{sect:bases}), but  only three are independent after integration by parts. Whenever possible, we choose the ones for which the covariant derivatives are contracted with the ``nearby'' tensor, so we can parametrize an action with the three independent kinetic terms
\begin{equation}
 S_0[g,\kappa] = \frac{1}{2} \int \sqrt{g} \left\{ \mathring{\nabla}_\lambda \kappa^\rho{}_{\mu\nu} \mathring{\nabla}^\lambda \kappa_\rho{}^{\mu\nu}
 + a_1 \mathring{\nabla}_\rho \kappa^\rho{}_{\mu\nu} \mathring{\nabla}_\lambda \kappa^\lambda{}^{\mu\nu}
 + a_2 \mathring{\nabla}_\rho \kappa^{\alpha\rho\beta} \mathring{\nabla}_\lambda \kappa_\alpha{}^\lambda{}_\beta \right\} \, .
\end{equation}
Obviously $S_0[g,\kappa]$ is not Weyl invariant in general, but we can express its infinitesimal Weyl transformation as
\begin{equation}
 \delta^W_\sigma S_0 = \int \sqrt{g} J^\lambda \partial_\lambda \sigma \, .
\end{equation}
The operation introduces a vector current $J^\lambda$ that reads
\begin{equation}
\begin{split}
 J^\lambda = & - \frac{d-2}{4} \mathring{\nabla}^\lambda (\kappa^2) + \left(1 + \frac{a_1(d+2)}{2} + \frac{a_2}{2}\right) \mathring{\nabla}_\rho \kappa^{\rho\mu\nu} \kappa^\lambda{}_{\mu\nu} - \kappa^{\rho\mu\nu} \mathring{\nabla}_\rho \kappa^\lambda{}_{\mu\nu} \\\nonumber
 & \quad +\left(2+\frac{a_2(d-2)}{2}\right) \mathring{\nabla}_\mu \kappa^{\rho\mu\nu} \kappa_\rho{}^\lambda{}_\nu - 2 \kappa^{\rho\mu\nu} \mathring{\nabla}_\mu \kappa_\rho{}^\lambda{}_\nu \, .
 \end{split}
\end{equation}
A necessary condition for Weyl invariance is that the current $J^\lambda$
can be expressed as the divergence of a symmetric tensor \cite{Erdmenger:1997wy}, so we impose 
that
\begin{equation}
 J^\lambda = \mathring{\nabla}_\rho j^{\rho\lambda} \, ,
\end{equation}
where $j^{\lambda\rho}=j^{\rho\lambda}$. In order for this to be possible,
we must constrain the two free coefficients of $S_0$ to be
\begin{equation}
 a_1 = - \frac{4(d-4)}{(d+2)(d-2)} \, ,
 \qquad
 a_2 = - \frac{8}{d-2} \, .
\end{equation} 
In this case the current $J^\lambda$ manifestly becomes a divergence
\begin{equation}
\begin{split}
J^\lambda &= - \frac{d-2}{4} \mathring{\nabla}^\lambda (\kappa^2) - \mathring{\nabla}_\rho (\kappa^{\lambda\mu\nu} \kappa^\rho{}_{\mu\nu}) - 2 \mathring{\nabla}_\rho (\kappa^{\mu\lambda\nu} \kappa_\mu{}^\rho{}_\nu) \, ,
\\
j^{\rho\lambda} &=
- \frac{d-2}{4} g^{\rho\lambda} \kappa^2 -  \kappa^{\lambda\mu\nu} \kappa^\rho{}_{\mu\nu} - 2 \kappa^{\mu\lambda\nu} \kappa_\mu{}^\rho{}_\nu\,,
\end{split}
\end{equation}
with the shorthand $\kappa^2= \kappa^{\mu\nu\rho}\kappa_{\mu\nu\rho}$.

We have not achieved conformal invariance yet. To do so, we must turn our attention to the couplings of $\kappa$ with the Ricci tensor and scalar curvature, which we choose to write in terms of the Schouten tensor and its trace. There is only one possible coupling of $\kappa$ with $\mathring{{\cal J}}$ and two independent couplings with the Schouten, that we parametrize as
\begin{equation}
 S_1[g,\kappa] = \int \sqrt{g} \left\{ a_3 \mathring{\mathcal{J}} \kappa^2 + a_4 \mathring{K}_{\mu\nu} \kappa^\mu{}_{\alpha\beta} \kappa^{\nu\alpha\beta} + a_5 \mathring{K}_{\mu\nu} \kappa_\alpha{}^\mu{}_\beta \kappa^{\alpha\nu\beta} \right\} \, .
\end{equation}
Using the known transformations of the Schouten tensor and of the scalar curvature, we combine the infinitesimal transformations of the actions $S_0$ and $S_1$ and find
\begin{equation}
 \delta^W_\sigma (S_0 + S_1) = \int \sqrt{g} \left\{ \left( \frac{d-2}{4} - a_3 \right) \kappa^2 \mathring{\square} \sigma + \left( 1 - a_4 \right) \kappa^{\lambda\mu\nu} \kappa^\rho{}_{\mu\nu} \mathring{\nabla}_\lambda \partial_\rho \sigma + \, \left( 2 - a_5 \right) \kappa^{\mu\rho\nu} \kappa_\mu{}^\lambda{}_\nu \mathring{\nabla}_\lambda \partial_\rho \sigma \right\} \, .
\end{equation}
It is then straightforward to choose the parameters of $S_1$ that cancel the combined transformation
\begin{equation}
a_3 = \frac{d-2}{4} \, ,
\qquad
a_4 = 1 \, ,
\qquad
a_5 = 2 \, .
\end{equation}

The actions $S_0$ and $S_1$ do not cover all possible terms, but rather leave out contractions of two $\kappa$s with the Weyl tensor defined in \eqref{eq:Riem-to-Weyl&Schouten}. Since the Weyl tensor transforms covariantly,
any such contraction is manifestly Weyl invariant as seen using dimensional analysis. There are two independent constractions of $\kappa$ with the Weyl tensor, which can be proven using the hook-symmetries \eqref{HAproperties}
and the Bianchi identities. We defer this algebraic analysis to the Appendix \ref{sect:bases}.
Therefore, the Weyl invariant action for a hook-antisymmetric traceless tensor on a $d$-dimensional Riemannian space is
\begin{align}\label{eq:Conf-Action-Kappa}
 S_{\rm c}[g,\kappa] = \int \sqrt{g} & \left\{
 \frac{1}{2} \mathring{\nabla}_\lambda \kappa^\rho{}_{\mu\nu} \mathring{\nabla}^\lambda \kappa_\rho{}^{\mu\nu}
 - \frac{2(d-4)}{(d+2)(d-2)} \mathring{\nabla}_\rho \kappa^\rho{}_{\mu\nu} \mathring{\nabla}_\lambda \kappa^\lambda{}^{\mu\nu}
 - \frac{4}{d-2} \mathring{\nabla}_\rho \kappa^{\alpha\rho\beta} \mathring{\nabla}_\lambda \kappa_\alpha{}^\lambda{}_\beta \right. \\\nonumber
 &  \left. 
 +  \frac{d-2}{4} \mathring{\mathcal{J}} \kappa^2 + \mathring{K}_{\mu\nu} \kappa^\mu{}_{\alpha\beta} \kappa^{\nu\alpha\beta} + 2 \mathring{K}_{\mu\nu} \kappa_\alpha{}^\mu{}_\beta \kappa^{\alpha\nu\beta} + \alpha_1 \mathring{W}^{\alpha\beta\mu\nu} \kappa^\lambda{}_{\alpha\beta} \kappa_{\lambda\mu\nu} + \alpha_2 \mathring{W}^{\alpha\mu\beta\nu} \kappa_\alpha{}^\lambda{}_\beta \kappa_{\mu\lambda\nu} \right\} \, ,
\end{align}
and the couplings $\alpha_1$ and $\alpha_2$ are arbitrary.
The presence of arbitrary couplings is expected based on the fact that $\kappa$
allows for manifestly invariant contractions with the Weyl tensor,
as opposed to the situation seen in the previous subsection with the vector.
The conformal actions of $p$-forms and traceless symmetric rank-$p$ tensors also each allow for a single interaction with the Weyl tensor \cite{Erdmenger:1997gy,Erdmenger:1997wy}, but we find that hook-antisymmetry allows for two parameters instead.\footnote{%
As further test, we have computed the action \eqref{eq:Conf-Action-Kappa} also by brute force using the most general ansatz and the Mathematica package \emph{xAct} and \emph{xTras} \cite{xact,Nutma:2013zea}.
We have checked that its variational energy-momentum tensor is
conserved and traceless, as expected from diffeomorphism and conformal invariances. We have also checked that in the flat-space limit the energy-momentum tensor is a primary operator, i.e., it is in the kernel of the action of the generators of special conformal transformations, using the procedure described in \cite{Osborn:2015rna,Stergiou:2022qqj}.
\label{footnote-checks}
}

In order to better understand the geometrical content of the conformal action \eqref{eq:Conf-Action-Kappa}, we now attempt a procedure similar to the one shown in the previous section in which some derivatives of the vector form field were replaced by their exterior differentiation.
The hook-antisymmetric tensor can be seen either as a vector of antisymmetric tensors, or as a vector-valued $2$-form tensor. As such, we have two natural ways to construct curvatures from exterior differentiation.
Let us introduce the following generalized field strengths \cite{Curtright:1980yj,Curtright:1980yk}
\begin{equation}
 \begin{split}
  F_{\mu\nu\rho\sigma} & \equiv \mathring{\nabla}_\mu \kappa_{\nu\rho\sigma} - \mathring{\nabla}_\nu \kappa_{\mu\rho\sigma} \, , \\
  G^\lambda{}_{\mu\rho\sigma} & \equiv \mathring{\nabla}_\mu \kappa^\lambda{}_{\rho\sigma} + \mathring{\nabla}_\rho \kappa^\lambda{}_{\sigma\mu} + \mathring{\nabla}_\sigma \kappa^\lambda{}_{\mu\rho} \, .
 \end{split}
\end{equation}
The first tensor has the same symmetries of the Riemann tensor in a torsionfull spacetime and can be interpreted as the curvature $2$-form of the vector part of $\kappa$.\footnote{%
Indeed, it is quite easy to show that it is actually equal to the part of the full Riemann tensor which is linear in the derivatives of the tensorial part of the torsion tensor. We return to the MAG interpretation later.
}
The second tensor is instead a vector-valued $3$-form, coming from exterior differentiation of the antisymmetric indices of $\kappa$.
In order to rewrite the action \eqref{eq:Conf-Action-Kappa} in terms of the generalized field strengths, we must choose a third basis element, similarly to the choice made with the divergence of the vector in the previous subsection,
and our pick is $ \mathring{\nabla}_\lambda \kappa^\lambda{}_{\mu\nu}$. Notice that we have $\mathring{\nabla}_\lambda \kappa^\lambda{}_{\mu\nu}=G^\lambda{}_{\lambda\mu\nu}$, thus the third kinetic term shall actually be written as the trace of the second one.

At this point we need just computing the square of the generalized field strengths. For the first one we have
\begin{equation}
\begin{split}
 \int \sqrt{g} \, F^{\mu\nu\rho\sigma} F_{\mu\nu\rho\sigma}
 & = \int \sqrt{g} \left\{ 2 \mathring{\nabla}_\lambda \kappa^\rho{}_{\mu\nu}\mathring{\nabla}^\lambda \kappa_\rho{}^{\mu\nu} - 2 \mathring{\nabla}_\lambda \kappa^\lambda{}_{\mu\nu} \mathring{\nabla}_\rho \kappa^{\rho\mu\nu} + 2 \mathring{\mathcal{J}} \kappa^2 \right. \\
 & \qquad \left.+ 2d \mathring{K}_{\alpha\beta} \kappa^\alpha{}_{\mu\nu} \kappa^{\beta\mu\nu} - 4 \mathring{W}_{\mu\alpha\nu\beta} \kappa^{\mu\rho\nu} \kappa^\alpha{}_\rho{}^\beta \right\} \, ,
 \end{split}
\end{equation}
up to integrations by parts. Similarly, for the second one we have
\begin{equation}
\begin{split}
 \int \sqrt{g} \, G_{\mu\nu\rho\sigma} G^{\mu\nu\rho\sigma}
 &
 = \int \sqrt{g} \left\{ 3 \mathring{\nabla}_\lambda \kappa^\rho{}_{\mu\nu} \mathring{\nabla}^\lambda \kappa_\rho{}^{\mu\nu}
 - 6 \mathring{\nabla}_\alpha \kappa^{\mu\alpha\nu} \mathring{\nabla}_\beta \kappa_\mu{}^\beta{}_\nu
 + 6 \mathring{\mathcal{J}} \kappa^2 + 6 \mathring{K}_{\alpha\beta} \kappa^\alpha{}_{\mu\nu} \kappa^{\beta\mu\nu} \right. \\
 &
 \qquad \left. + 6 (d-4) \mathring{K}_{\alpha\beta} \kappa^{\mu\alpha\nu} \kappa_\mu{}^\beta{}_\nu - 6 \mathring{W}_{\alpha\mu\beta\nu} \kappa^{\alpha\rho\beta} \kappa^\mu{}_\rho{}^\nu - 3 \mathring{W}_{\alpha\beta\mu\nu} \kappa_\lambda{}^{\alpha\beta} \kappa^{\lambda\mu\nu} \right\} \, .
 \end{split}
\end{equation}
Notice that in the expansion of the generalized field strengths we have also exploited the definition given in Eq.~\eqref{eq:Riem-to-Weyl&Schouten} to express them is the same basis adopted in Eq.~\eqref{eq:Conf-Action-Kappa}. The final step is to rewrite the conformal action for the hook-antisymmetric traceless tensor $\kappa$ as
\begin{equation}\label{eq:Conf-Action-Kappa-2}
 \begin{split}
 S_{\rm c}[g,\kappa] &= \int \sqrt{g} \left\{ \frac{d-6}{4(d-2)} F^2 + \frac{2}{3(d-2)} G^2 + \frac{d^2-8d+4}{2(d+2)(d-2)} \left(G^\lambda{}_{\lambda\mu\nu} \right)^2 + \frac{d(d-6)}{4(d-2)} \mathring{\mathcal{J}} \kappa^2 - \frac{d-6}{2} \mathring{K}_{\alpha\beta} \kappa^\alpha{}_{\mu\nu} \kappa^{\beta\mu\nu} \right. \\
 & \qquad\qquad \left.  - \frac{2(d-6)}{d-2} \mathring{K}_{\alpha\beta} \kappa_\mu{}^\alpha{}_\nu \kappa^{\mu\beta\nu}  + \alpha'_1 \mathring{W}_{\alpha\beta\mu\nu} \kappa^{\lambda\alpha\beta} \kappa_\lambda{}^{\mu\nu}
 + \alpha'_2 \mathring{W}_{\mu\alpha\nu\beta} \kappa^{\mu\rho\nu} \kappa^\alpha{}_\rho{}^\beta
 \right\} \, ,
\end{split}
\end{equation}
where a simple shift in the free coupling constants has occurred
\begin{equation}
 \alpha'_1 = \alpha_1 + \frac{2}{d-2} \, , \qquad \alpha'_2 = \alpha_2 + 1 \, .
\end{equation}

\subsection*{Intermezzo: the limit $d=6$ in flat space of the conformal traceless hook-antisymmetric action}

The form of the action \eqref{eq:Conf-Action-Kappa-2} suggests that something might happen in the limit $d=6$, similarly to the limit $d=4$ of the conformal vector's action that enjoys an Abelian gauge invariance. In fact, specializing to $d=6$ and choosing $\alpha'_1=\alpha'_2=0$, the action becomes
\begin{equation}
 S_{\rm c}[g,\kappa] = \frac{1}{6} \int \sqrt{g} \left\{ G^2 - \frac{3}{4} (G^\lambda{}_{\lambda\mu\nu})^2 \right\} \, .
\end{equation}
Even though this action has a remarkably simple form, in the flat-space limit it does not have the full enhanced set of symmetries discussed by Curtright \cite{Curtright:1980yj} (the relative coefficient should be $-3$ instead of $-\frac{3}{4}$, and there is no natural spacetime dimension $d$ for which this is true, we already checked). However, using the fact that the $G$ curvature \emph{is} invariant under the flat-space transformation
\begin{equation}
 \kappa^\lambda{}_{\mu\nu} \rightarrow \kappa^\lambda{}_{\mu\nu} + \partial_\mu S^\lambda{}_\nu - \partial_\nu S^\lambda{}_\mu \, ,
\end{equation}
parametrized by the symmetric trace-free tensor $S^\lambda{}_\mu$, which is one of the Curtright's transformations \cite{Curtright:1980yj}, we deduce that the conformal action \eqref{eq:Conf-Action-Kappa-2} has at least one new gauge invariance in $d=6$ in the flat-space limit. Therefore we conclude that an analogy with the conformal vectors in $d=4$ arises when we study hook anti-symmetric trace-free tensors in $d=6$. However, this analogy is much weaker because it is realized only in the flat-space limit.

Before moving on, notice that, always in the flat space-limit, the $G$-curvature is also invariant under the transformation
\begin{equation}
 \kappa^\lambda{}_{\mu\nu} \rightarrow \kappa^\lambda{}_{\mu\nu} + \partial^\lambda \partial_\mu \zeta_\nu - \partial^\lambda \partial_\nu \zeta_\mu \, ,
\end{equation}
where $\zeta_\mu$ is a transverse $1$-form. This transformation is a special case of the previous one, since we can take $S_{\mu\nu}=\partial_\mu \zeta_\nu + \partial_\nu \zeta_\mu$, provided $\zeta_\mu$ is transverse. Therefore, the mode which is parametrized by a $2$-form, say $A_{[\mu\nu]}$, is the only longitudinal mode that actually appears in the spin-parity decomposition of the conformal action in the flat-space limit in six spacetime dimensions (we return on the analysis of the modes in Sect.~\ref{sect:MaptoMags}).

\subsection*{The conformal traceless hook-antisymmetric operator}

We conclude this section by presenting the conformally covariant operator acting on $\kappa$. From $S_{\rm c}[g,\kappa]$ it is straightforward to obtain the associated rank two operator, ${\Delta_{\kappa}}$, that acts on traceless hook-antisymmetric tensors
\begin{equation}
 \begin{split}
  (\Delta_{\kappa}& \cdot \kappa)^\mu{}_{\alpha\beta} 
 =
 g^{\mu\rho}\,g_{\alpha\nu}\,g_{\beta\sigma}\frac{1}{\sqrt{g}}\frac{\delta S_{\rm c}[g,\kappa]}{\delta \kappa^\rho{}_{\nu\sigma} }
 =
 {({\cal P}_{\kappa})^\mu{}_{\alpha\beta}}^\rho{}_{\nu\sigma}\Bigg\{ - \mathring{\square}\,\kappa_\rho{}^{\nu\sigma} 
 + \frac{4(d-4)}{(d+2)(d-2)}\mathring{\nabla}_\rho \mathring{\nabla}_\gamma  \kappa^{\gamma\nu\sigma} +\\
& -\frac{8}{d-2}  \mathring{\nabla}^{[\nu }   \mathring{\nabla}_\gamma        \kappa_\rho{}^{\sigma]\gamma} 
+ \frac{d-2}{2}\kappa_\rho{}^{\nu\sigma}\mathring{\mathcal{J}} 
+ 2\kappa^{\gamma\nu\sigma} \mathring{K}_{\rho\gamma} 
+ 4\kappa_\rho{}^{[\nu|\gamma|} \mathring{K}^{\sigma]}{}_{\gamma}  
+2\alpha_1 \kappa_\rho{}^{\gamma\epsilon} \mathring{W}^{\nu\sigma}{}_{\gamma\epsilon}
+ 2\alpha_2 \kappa^{\gamma[\nu|\epsilon|} \mathring{W}_{\rho\gamma}{}^{\sigma]}{}_{\epsilon}  \Bigg\},
 \end{split}
\end{equation}
where $({\cal P}_{\kappa})$ is the unique projector on the space of traceless hook antisymmetric tensors. The projector $({\cal P}_{\kappa})$ is defined by the properties
\begin{align}
&
{({\cal P}_{\kappa})^\mu{}_{\alpha\beta}}^\rho{}_{\nu\sigma}
={({\cal P}_{\kappa})^\mu{}_{[\alpha\beta]}}^\rho{}_{\nu\sigma}\,,
&&
{({\cal P}_{\kappa})^\mu{}_{\mu\beta}}^\rho{}_{\nu\sigma}=0\,,
\nonumber
\\
&
{({\cal P}_{\kappa})^\mu{}_{\alpha\beta}}^\rho{}_{\nu\sigma}={({\cal P}_{\kappa})^\rho{}_{\nu\sigma}}^\mu{}_{\alpha\beta} \,,
&&
{({\cal P}_{\kappa})^\mu{}_{\alpha\beta}}^\rho{}_{\nu\sigma}
{({\cal P}_{\kappa})_\rho{}^{\nu\sigma}}^\gamma{}_{\lambda\theta}
={({\cal P}_{\kappa})^\mu{}_{\alpha\beta}}^\gamma{}_{\lambda\theta}
\,, \\
&
{({\cal P}_{\kappa})_{\mu\alpha\beta}}^\rho{}_{\nu\sigma}
+{({\cal P}_{\kappa})_{\alpha\beta\mu}}^\rho{}_{\nu\sigma}
+{({\cal P}_{\kappa})_{\beta\mu\alpha}}^\rho{}_{\nu\sigma}=0
\,.
&&
\nonumber
\end{align}
Its explicit form is
\begin{equation}
{({\cal P}_{\kappa})^\mu{}_{\alpha\beta}}^\rho{}_{\nu\sigma}
=
-\frac{1}{6}\,\delta_{\beta}{}^{\rho}\,\delta_{\nu}{}^{\mu}\,g_{\alpha\sigma} 
+ \frac{1}{2(d-1)}\,\delta_{\beta}{}^{\mu}\,\delta_{\nu}{}^{\rho}\,g_{\alpha\sigma}
+\frac{1}{6}\,g^{\mu\rho}g_{\alpha\nu}g_{\beta\sigma} 
- (\alpha\leftrightarrow\beta)-(\nu\leftrightarrow\sigma)
+ (\alpha\leftrightarrow\beta,\, \nu\leftrightarrow\sigma)\,,
\end{equation}
from which one can check that ${({\cal P}_{\kappa})^\mu{}_{\alpha\beta}}_\mu{}^{\alpha\beta}=\frac{1}{3}d(d+2)(d-2)$.
From the definition one can see that each ``triplet'' of indices in the projector has precisely the same symmetry properties of $\kappa$ (including cyclicity) so that it acts like the identity in the space of hook antisymmetric tensors. Of course, it is also symmetric under the interchange of the first triplet of indices with the second. We wrote the operator using such a projector to keep its expression as simple and compact as possible.

\subsection{Conformal coupling of the hook-symmetric traceless tensor} \label{sect:hook-symmetric}

For the construction of the conformal action of the traceless hook-symmetric tensor $\psi^\mu{}_{\nu\rho}$ with the properties \eqref{HSproperties} we can follow closely the same steps of the derivation of the previous subsection.
Similarly to the previous case there are five terms that can be constructed with derivatives of $\psi$ of the form $(\mathring{\nabla}\psi)^2$ and three are independent up to integrations by parts. We define
\begin{equation}
S_0[g,\psi]=\frac{1}{2} \int d^d x\sqrt{g} \left\{ \mathring{\nabla}_\lambda \psi^\mu{}_{\nu\xi}\, \mathring{\nabla}^\lambda \psi_\mu{}^{\nu\xi} + a_1 \mathring{\nabla}_\sigma \psi^\sigma{}_{\nu\xi} \,\mathring{\nabla}_\alpha \psi^\alpha{}^{\nu\xi} + a_2 \mathring{\nabla}_\sigma \psi^{\alpha\sigma\beta}\, \mathring{\nabla}_\gamma \psi_\alpha{}^\gamma{}_\beta \right\}\,.
\end{equation}
The Weyl weight of $\psi$ is $w_{\psi}=\frac{4-d}{2}$ as before, so we only need
\begin{equation}
\delta^W_\sigma \psi^\mu{}_{\nu\rho} = -\frac{d-4}{2}\psi^\mu{}_{\nu\rho}\,,
\end{equation}
besides the transformations already given in the previous subsection.

It is a simple computation to derive the Weyl transformation of covariant derivatives of the tensor
\begin{equation}
\begin{split}
 \delta_\sigma \mathring{\nabla}_\lambda \psi^\mu{}_{\nu\xi} &=  \frac{2-d}{2} \psi^\mu{}_{\nu\xi} \partial_\lambda \sigma + \delta^\mu{}_\lambda  \partial_\gamma \sigma \psi^\gamma{}_{\nu\xi} - \partial^\mu \sigma \psi_{\lambda\nu\xi}  - \partial_\nu \sigma  \psi^\mu{}_{\lambda\xi}
 + g_{\nu\lambda} \partial_\delta \sigma  \psi^{\mu\delta}{}_\xi  - \partial_\xi \sigma\psi^\mu{}_{\nu\lambda}  + g_{\lambda\xi}\partial^\alpha \sigma\psi^\mu{}_{\nu\alpha}\,,
\end{split}
\end{equation}
We write the transformation of $S_0$ as
\begin{equation}
\delta_\sigma S_0 = \int d^d x\sqrt{g} \,\partial_\mu \sigma J^\mu \,,
\end{equation}
which implicitly defines $J^\mu$. It can be explicitly computed as
\begin{equation}
\begin{split}
\delta_\sigma S_0 = \int d^d x\sqrt{g} \, \Bigg\{  & \frac{2-d}{4} \mathring{\nabla}^\mu (\psi^2)   - 2\,\psi^\xi{}_{\nu\lambda}\mathring{\nabla}^\lambda  \psi_\xi{}^{\nu\mu} + \left(2 + \frac{a_2(2+d)}{2} \right)\psi^{\gamma\mu}{}_{\xi} \mathring{\nabla}_\lambda  \psi_\gamma{}^{\lambda\xi}  \\
&    - \psi_{\lambda\nu\xi}  \mathring{\nabla}^\lambda\, \psi^{\mu\nu\xi}   + \frac{1}{2} \left( a_1(d-2) - a_2 +2\right)  \psi^\mu{}_{\nu\xi}\ \mathring{\nabla}_\alpha\psi^{\alpha\nu\xi}        \Bigg\} \partial_\mu \sigma\,.
\label{S0_var}
\end{split}
\end{equation}
As condition for Weyl invariance we must have $J^\mu= \mathring{\nabla}_\xi j^{\mu\xi}$.
This is possible only if we chose
\begin{equation}
 a_1 = -4 \,\frac{4+d}{(d-2)(d+2)} \, ,  \qquad a_2 = - \frac{8}{2+d}\,,
\end{equation} 
In this way we obtain
\begin{equation}
\begin{split}
J^\lambda &=
-\frac{2-d}{4}\mathring{\nabla}^\lambda (\psi^2) + 2 \mathring{\nabla}_\mu(\psi_\xi{}^\lambda{}_\nu \psi^{\xi\mu\nu}) + \mathring{\nabla}_\mu(\psi^\lambda{}_{\nu\xi}\psi^{\mu\nu\xi})\,,
\\
j^{\mu\lambda} &=-\frac{2-d}{4}\psi^2\,g^{\mu\lambda} + 2 \psi_\xi{}^\lambda{}_\nu \psi^{\xi\mu\nu} + \psi^\lambda{}_{\nu\xi}\psi^{\mu\nu\xi}
\,.
\end{split}
\end{equation}
The curvature dependent terms must be chosen so that the Weyl transformation of $S_0$ is balanced. Using only the Schouten tensor and its trace
we parametrize them as 
\begin{equation}
 S_1[g,\psi] = \int d^d x\sqrt{g} \left\{ a_3 \mathring{\mathcal{J}} \,\psi^2 + a_4 \mathring{K}_{\mu\nu} \,\psi^{\mu\gamma\sigma} \psi^\nu{}_{\gamma\sigma} + a_5 \mathring{K}_{\mu\nu}\,\psi^{\alpha\mu\beta}\psi_\alpha{}^\nu{}_\beta \right\}\,.
\end{equation}
The combined transformation of the two actions is
\begin{equation}
\begin{split}
 \delta_\sigma (S_0 + S_1) &= \int d^d x\sqrt{g}\,  \left\{ -\left(a_3 + \frac{2-d}{4}\right) \psi^2\, \mathring{\square} \sigma + \left( 1-a_4\right) \psi^\lambda{}_{\nu\xi}\psi^{\mu\nu\xi} \mathring{\nabla}_\lambda \partial_\mu \sigma
  +  \left(2-a_5 \right)\,\psi_\xi{}^\lambda{}_\nu \psi^{\xi\mu\nu}\mathring{\nabla}_\lambda \partial_\mu \sigma \right\}\,,
\end{split}
\end{equation}
so the requirement $\delta_\sigma (S_0 + S_1) = 0$ gives
\begin{equation}
a_3 =\frac{d-2}{4}\,,\qquad
a_4=1 \,,\qquad
a_5=2\,. 
\end{equation}
Much like in the previous subsection, there still are two independent interactions with the Weyl curvature tensor that are manifestly invariant.
In conclusion we have the following conformally invariant action
\begin{equation}\label{eq:conf-action-psi}
\begin{split}
 S_{\rm conf}[g,\psi] &=  \int d^d x\sqrt{g} \left\{
 \frac{1}{2} \mathring{\nabla}_\lambda \psi^\mu{}_{\nu\xi} \mathring{\nabla}^\lambda \psi_\mu{}^{\nu\xi} 
 -2 \frac{4+d}{(d-2)(d+2)} \mathring{\nabla}_\sigma \psi^\sigma{}_{\nu\xi} \mathring{\nabla}_\alpha \psi^\alpha{}^{\nu\xi}
 - \frac{4}{2+d} \mathring{\nabla}_\sigma \psi^{\alpha\sigma\beta} \mathring{\nabla}_\gamma \psi_\alpha{}^\gamma{}_\beta \right. \\
 & \quad \left. 
 + \frac{d-2}{4}  \mathring{\mathcal{J}} \psi^2
 +  \mathring{K}_{\mu\nu} \psi^{\mu\gamma\sigma} \psi^\nu{}_{\gamma\sigma}
 + 2  \mathring{K}_{\mu\nu}\psi^{\alpha\mu\beta}\psi_\alpha{}^\nu{}_\beta
 + \beta_1 \mathring{W}_{\alpha\beta\mu\nu} \psi^{\lambda\alpha\mu} \psi_\lambda{}^{\beta\nu} 
 + \beta_2 \mathring{W}_{\alpha\mu\beta\nu} \psi^{\alpha\lambda\mu}\psi^\beta{}_\lambda{}^\nu\right\} \,,
\end{split}
\end{equation}
that depends only on two parameters as with the case of the hook-antisymmetric tensor.\footnote{%
We have performed the same checks as the hook-antisymmetric case described in footnote \ref{footnote-checks}.
}

\subsection*{The conformal traceless hook-symmetric operator}

In strict analogy with the previous subsection, we conclude by presenting the conformally covariant operator acting on $\psi$. Of course, from $S_{\rm conf}[g,\psi]$ we obtain such a rank two operator, ${\Delta_{\psi}}$, that acts on traceless hook-antisymmetric tensors, by functional differentiation
\begin{equation}
 \begin{split}
 (\Delta_{\psi}& \cdot \psi)^\mu{}_{\alpha\beta} 
  =
 g^{\mu\rho}\,g_{\alpha\nu}\,g_{\beta\sigma}\frac{1}{\sqrt{g}}\frac{\delta S_{\rm conf}[g,\psi]}{\delta \psi^\rho{}_{\nu\sigma} }
  =
 {({\cal P}_{\psi})^\mu{}_{\alpha\beta}}^\rho{}_{\nu\sigma}\Bigg\{ - \mathring{\square}\,\psi_\rho{}^{\nu\sigma} + \frac{4(d+4)}{(d+2)(d-2)}\mathring{\nabla}_\rho \mathring{\nabla}_\gamma  \psi^{\gamma\nu\sigma} +\\
& + \frac{8}{d-2}  \mathring{\nabla}^{(\nu }   \mathring{\nabla}_\gamma       \psi_\rho{}^{\sigma)\gamma} 
+ \frac{d-2}{2}\psi_\rho{}^{\nu\sigma}\mathring{\mathcal{J}} 
+ 2\psi^{\gamma\nu\sigma} \mathring{K}_{\rho\gamma} 
+ 4\psi_\rho{}^{(\nu|\gamma|} \mathring{K}^{\sigma)}{}_{\gamma} 
+ 4\beta_1 \psi_\rho{}^{\gamma\epsilon} \mathring{W}^{(\nu}{}_{\gamma}{}^{\sigma)}{}_{\epsilon}
+ 2\beta_2 \psi^{\gamma(\nu|\epsilon|} \mathring{W}_{\rho\gamma}{}^{\sigma)}{}_{\epsilon}  \Bigg\},
 \end{split}
\end{equation}
where now $({\cal P}_{\psi})$ is the unique projector on the space of traceless hook-symmetric tensors defined by the properties
\begin{align}
&
{({\cal P}_{\psi})^\mu{}_{\alpha\beta}}^\rho{}_{\nu\sigma}
={({\cal P}_{\psi})^\mu{}_{(\alpha\beta)}}^\rho{}_{\nu\sigma}\,,
&&
{({\cal P}_{\psi})^\mu{}_{\mu\beta}}^\rho{}_{\nu\sigma}=
\delta^\alpha{}_\mu{({\cal P}_{\psi})^\mu{}_{\alpha\beta}}^\rho{}_{\nu\sigma}
=0\,,
\nonumber
\\
&
{({\cal P}_{\psi})^\mu{}_{\alpha\beta}}^\rho{}_{\nu\sigma}={({\cal P}_{\psi})^\rho{}_{\nu\sigma}}^\mu{}_{\alpha\beta} \,,
&&
{({\cal P}_{\psi})^\mu{}_{\alpha\beta}}^\rho{}_{\nu\sigma}
{({\cal P}_{\psi})_\rho{}^{\nu\sigma}}^\gamma{}_{\lambda\theta}
={({\cal P}_{\psi})^\mu{}_{\alpha\beta}}^\gamma{}_{\lambda\theta}
\,, \\
&
{({\cal P}_{\psi})_{\mu\alpha\beta}}^\rho{}_{\nu\sigma}
+{({\cal P}_{\psi})_{\alpha\beta\mu}}^\rho{}_{\nu\sigma}
+{({\cal P}_{\psi})_{\beta\mu\alpha}}^\rho{}_{\nu\sigma}=0
\,.
&&
\nonumber
\end{align}
The explicit form of $({\cal P}_{\psi})$ is
\begin{equation}
 \begin{split}
{({\cal P}_{\psi})^\mu{}_{\alpha\beta}}^\rho{}_{\nu\sigma}
&=
-\frac{1}{6}\,\delta_{\beta}{}^{\rho}\,\delta_{\nu}{}^{\mu}\,g_{\alpha\sigma}
- \frac{1}{6(d-1)}\,\delta_{\beta}{}^{\mu}\,\delta_{\nu}{}^{\rho}\,g_{\sigma\alpha}
+\frac{1}{6}\,g^{\mu\rho}g_{\alpha\nu}g_{\beta\sigma} 
+ (\alpha\leftrightarrow\beta) 
+ (\nu\leftrightarrow\sigma)
+ (\alpha\leftrightarrow\beta,\, \nu\leftrightarrow\sigma)\\
&+ \frac{1}{3(d-1)}\,\delta_{\nu}{}^{\mu}\,\delta_{\sigma}{}^{\rho}\,g_{\alpha\beta}  
+ (\nu\leftrightarrow\sigma)
+ \frac{1}{3(d-1)}\,\delta_{\beta}{}^{\mu}\,\delta_{\alpha}{}^{\rho}\,g_{\nu\sigma}  
+ (\alpha\leftrightarrow\beta) 
- \frac{2}{3(d-1)}g^{\mu\rho}g_{\alpha\beta}g_{\nu\sigma}
\,,
 \end{split}
\end{equation}
from which one can check that ${({\cal P}_{\psi})^\mu{}_{\alpha\beta}}_\mu{}^{\alpha\beta}=\frac{1}{3}d(d+2)(d-2)$.
The same considerations on the symmetry properties done for
$({{\cal P}}_{\kappa})$
in the previous section apply here replacing $\kappa$ with $\psi$.

\section{Metric-affine interpretation: hook-antisymmetric}
\label{sect:MaptoMags}

In order to discuss the interpretation of the conformal action given in Sect.~\ref{sect:hook-antisymmetric} as a MAG theory we follow the procedure conveniently discussed by Baldazzi et al.\ in Ref.~\cite{Baldazzi:2021kaf}.
We must specialize to the case $d=4$ in which $\kappa$ can be naturally interpreted as the purely tensorial part of the torsion as in the decomposition \eqref{eq:T-Q-decomposition}.

In practice, we consider a MAG in which all connection related tensors, except for $\kappa$, are zero, so only the contortion part is active in \eqref{eq:distortion-decomposition}. Using \eqref{eq:contortion-def} and $T^\mu{}_{\nu\rho}=\kappa^\mu{}_{\nu\rho}$, we see that
\begin{equation}
\begin{split}
 K^\mu{}_{\nu\rho}= \frac{1}{2}\Bigl(\kappa_\rho{}^\mu{}_\nu+\kappa_\nu{}^\mu{}_\rho-\kappa^\mu{}_{\nu\rho}\Bigr)
 = \kappa_\rho{}^\mu{}_\nu\,,
\end{split}
\end{equation}
thanks to the cyclicity property of $\kappa$ in \eqref{HAproperties}. Thus, the covariant derivative is particularly simple
\begin{equation}\label{eq:kappa-mag-covd}
\begin{split}
 \nabla_\mu v^\nu = \mathring{\nabla}_\mu v^\nu + \kappa_\mu{}^\nu{}_\rho v^\rho\,,
\end{split}
\end{equation}
where $v^\mu$ is an arbitrary vector. By construction this covariant derivative is compatible, but it has the truly tensorial torsion $\kappa$.
Below we concentrate on two main aspects: the analysis of the degrees of freedom
introduced by $\kappa$ from the point of view of the Einstein representation of a MAG (in which $\kappa$ is an additional field coupled to the Riemannian geometry based on $g_{\mu\nu}$ and $\mathring{\nabla}$),
and the construction of a MAG theory in the Cartan representation (i.e., with curvatures of $\nabla$) that coincides with \eqref{eq:Conf-Action-Kappa} at the leading interaction order and thus can be seen as an interacting perturbation of the conformal action.

\subsection{Embedding into MAG: hook-antisymmetric degrees of freedom}\label{sect:MaptoMags-HA}

We specialize \eqref{eq:Conf-Action-Kappa} to $d=4$, but also couple it with a scalar field $\phi$ with canonical weight $w_\phi=(2-d)/2=-1$ for generality. The scalar field can be used, for example, to break conformal invariance by acquiring a vev $\phi_0$ \cite{Ghilencea:2018thl}.
We thus consider the following action
\begin{align}
  S_I [g,\kappa,\phi] = \int \sqrt{g} & \left\{
  \frac{1}{2} \mathring{\nabla}_\lambda \kappa^\rho{}_{\mu\nu} \mathring{\nabla}^\lambda \kappa_\rho{}^{\mu\nu}
  - 2 \mathring{\nabla}_\rho \kappa^{\alpha\rho\beta} \mathring{\nabla}_\lambda \kappa_\alpha{}^\lambda{}_\beta
  + \frac{1}{2} \mathring{\mathcal{J}} \kappa^2
  + \mathring{K}_{\mu\nu} \kappa^\mu{}_{\alpha\beta} \kappa^{\nu\alpha\beta} + 2 \mathring{K}_{\mu\nu} \kappa_\alpha{}^\mu{}_\beta \kappa^{\alpha\nu\beta} \right. \\\nonumber
  & \, \left. \, +  \alpha_1 \mathring{W}^{\alpha\beta\mu\nu} \kappa^\lambda{}_{\alpha\beta} \kappa_{\lambda\mu\nu} + \alpha_2 \mathring{W}^{\alpha\mu\beta\nu} \kappa_\alpha{}^\lambda{}_\beta \kappa_{\mu\lambda\nu} + \frac{1}{2} \left( \partial_\mu \phi \right)^2 + \frac{1}{12} \phi^2 \mathring{R} - \frac{\lambda}{4!} \phi^4 - \frac{\zeta}{2} \phi^2 \kappa^2 + \frac{\alpha_3}{2} \mathring{W}^2_{\mu\nu\rho\sigma} \right\} \, ,
\end{align}
which has also an additional Weyl-squared term, $\mathring{W}^2_{\mu\nu\rho\sigma}=\mathring{W}_{\mu\nu\rho\sigma}\mathring{W}^{\mu\nu\rho\sigma}$, that is also manifestly conformal invariant.
It is easy to see that this is the most general (classically) Weyl invariant action in four dimensions, which is at most quadratic in $\kappa$ and which depends also on the metric and the scalar field $\phi$.

The general action can be expanded quadratically in flat space, requiring $\phi=\phi_0$ (constant) and diagonalized in the spin components using the spin-projectors \cite{Percacci:2020ddy} and results of \cite{Baldazzi:2021kaf} in the so-called Einstein basis. We denote $q^2=q_\mu q^\mu$ the square momentum and recognize the momentum-space kinetic terms in the flat space limit for all the propagating degrees of freedom
\begin{align} \label{eq:modes-antisymm}
 & a(2^+)_{gg} = - q^4 \frac{\alpha_3}{4} - \frac{q^2 \phi_0^2}{32}  \, , 
 && a(2^+)_{\kappa\kappa} = \frac{q^2}{2} - \zeta \phi_0^2 \, ,
 \nonumber\\
 & a(2^-)_{\kappa\kappa} = - \frac{q^2}{2} - \zeta \phi_0^2 \, , 
 && a(1^+)_{\kappa\kappa} = - \frac{q^2}{6} - \zeta \phi_0^2 \, , \\
 & a(1^-)_{\kappa\kappa} = \frac{q^2}{2} - \zeta \phi_0^2 \, , 
 && a(0^+)_{gg} = q^4 \frac{4 \alpha_3}{3} + \frac{q^2 \phi_0^2}{24} \, .
 \nonumber
\end{align}
The notation is as follows: for each matrix element $a$, the argument denotes the spin-parity content, while the subscripts indicate the fields that are responsible for its propagation.
Notice that there is no mixing between the tensor $\kappa$ of the torsion and the metric. In fact, even in the $2^+$ sector which could be sensible to mixing, we cannot find any because of the absence of conformally invariant terms of the form $\kappa \mathring{\nabla} \mathring{R}$. In contrast, the absence of mixing in the $1^-$ sector is a general consequence of the infinitesimal transformation of the metric tensor under diffeomorphisms, see \cite{Baldazzi:2021kaf}. Note also that, for a slightly more general analysis, one should also take into account the contributions to the $0^+$ sector from the scalar field $\phi$,
which, instead, are expected to mix with the rest.

The propagating modes in Eq.~\eqref{eq:modes-antisymm} appear in various forms.
There is the potential for both kinematical ghosts and nonunitary contributions,
as can be seen from the various possible signs and powers of $q^2$. A rather general discussion of the properties of the spin-decomposition of higher derivative gravity can be found in Ref.~\cite{percacci-book}.
Here we concentrate on the contributions coming from the hook-antisymmetric field $\kappa$, from which we evince that there are always two kinematical ghost modes, with spins $2^+$ and $1^-$ (so both spins with the ``correct'' parity),
that can be either massive, massless or tachyonic depending on the
value of the parameter $\zeta$ (assuming a nonzero vev $\phi_0$).

\subsection{Cartan's action for the hook-antisymmetric traceless tensor}

In this subsection we construct a ``geometric'' MAG action containing the Weyl-invariant action for the traceless hook-antisymmetric tensor in $d=4$ given in \eqref{eq:Conf-Action-Kappa}. The idea is to find an action that is constructed from the MAG curvatures and that, if expanded quadratically in the distortion/contortion, reduces to the one given in \eqref{eq:Conf-Action-Kappa}.
To this end, we start by considering the metric-compatible torsionful connection given in Eq.~\eqref{eq:kappa-mag-covd}.
Following the discussion of Ref.~\cite{Percacci:2020ddy}, we write down the most general action containing curvature squared terms.
Taking into account symmetry properties of the curvatures tensor, it is not difficult to prove that there are three independent contributions from the Riemann tensor squared, and two from the the Ricci tensor squared
(in general the Ricci tensor is not symmetric, $R_{\mu\nu}=R^{\rho}{}_{\mu\rho\nu}$).

However, by a direct computation, we can see that this prescribed action is not sufficient for our purpose, since the contribution proportional to $\mathring{\mathcal{J}}\kappa^2$ of \eqref{eq:Conf-Action-Kappa} would not be produced. It is therefore necessary to add at least one additional term,
which we choose with the schematic form $RT^2$ ($R^2$ is not independent from the rest of the prescribed action, because it is related to Riemann and Ricci squared contributions through the Gauss-Bonnet term \cite{Percacci:2020ddy}).
Using the above logic, we find the action
\begin{align}
 S_{\rm MAG}[g,\kappa] = \int \sqrt{g}   \Big\{ p\,R\,T^{\alpha\beta\gamma}\,&T_{\alpha\beta\gamma} + q\, R^{\mu\nu\rho\sigma}R_{\mu\nu\rho\sigma} + r\, R^{\mu\nu\rho\sigma}R_{\rho\sigma\mu\nu}  + s\, R^{\mu\nu\rho\sigma}R_{\mu\rho\nu\sigma}
+ t\, R^{\mu\nu}R_{\mu\nu} + u\, R^{\mu\nu}R_{\nu\mu}       \Big\}.
\end{align}
This action can be rewritten in the Einstein form imposing that $T^\mu{}_{\nu\rho} =\kappa^\mu{}_{\nu\rho}$. We fix the couplings by imposing that the contributions quadratic in $\kappa$ must reproduce the Weyl-invariant action for the traceless hook-antisymmetric tensor \eqref{eq:Conf-Action-Kappa}.
The explicit solution for the matching is
\begin{align} 
 & p=-\frac{19}{144} \, , 
 && q=-\frac{11}{12}\, ,
 &&  r=\frac{1}{4}\, ,
 && s=-\frac{4}{3}\,
 \nonumber\\
 & t=-\frac{1}{6} \, , 
 && u=-\frac{1}{4} \,  
 && \alpha_1=-\frac{7}{24} \, ,
 && \alpha_2=-\frac{19}{12}  \, . 
 \nonumber
\end{align}
Interestingly, the matching procedure fixes the couplings $\alpha_1, \alpha_2$ with the Weyl tensor, which had not previously been fixed by requiring Weyl invariance.
We have thus found a geometrical MAG action in the Cartan form that can be expanded as
\begin{equation}
 S_{\rm MAG}[g,\kappa] = S_{\rm conf}[g,\kappa] + S_{\rm int}\,,
\end{equation}
where the terms $S_{\rm int}$ are higher order interactions in the tensor $\kappa$. By construction, $S_{\rm int}$ contains (several) cubic and quartic
interactions of the torsion in the schematic forms $\kappa^2\partial \kappa$ and $\kappa^4$. We argue that the conformal action $S_{\rm conf}$ should provide a natural expansion point for the quantization of the field $\kappa$ and that the interactions might be treated perturbatively.
This could prove useful in the construction of a quantum theory for the MAG, which is a step that we hope to undertake in the future.

\section{Metric-affine interpretation: hook-symmetric}
\label{sect:MaptoMags-2}

Similarly to the previous section, we want to interpret the conformal action 
in Sect.~\ref{sect:hook-symmetric} as a MAG theory in $d=4$. In this case $\psi$ is interpreted as a purely tensorial part of the nonmetricity, and all other affine tensors are zero.
We have that only the disformation part is active in \eqref{eq:distortion-decomposition}. Using \eqref{eq:disformation-def} and $Q^\mu{}_{\nu\rho}=\psi^\mu{}_{\nu\rho}$, we see that
\begin{equation}
\begin{split}
 L^\mu{}_{\nu\rho} = \frac{1}{2}\Bigl(\psi_\rho{}^\mu{}_\nu+\psi_\nu{}^\mu{}_\rho-\psi^\mu{}_{\nu\rho}\Bigr)
 = 2\psi_{(\rho}{}^\mu{}_{\nu)}\,,
\end{split}
\end{equation}
thanks to the cyclicity property of $\psi$ in \eqref{HSproperties}.
The covariant derivative is almost as simple as the one of the previous section
\begin{equation}\label{eq:connection-psi}
\begin{split}
 \nabla_\mu v^\nu = \mathring{\nabla}_\mu v^\nu + \psi_\mu{}^\nu{}_\rho v^\rho+\psi_\rho{}^\nu{}_\mu v^\rho\,,
\end{split}
\end{equation}
where $v^\mu$ is an arbitrary vector. By construction it is torsionless, but not metric. For the remainder of this section we follow similar steps to the previous one, that is, we discuss the spin degrees of freedom of the action \eqref{eq:conf-action-psi} in the Einstein representation and then we construct
an interacting MAG which reduces to \eqref{eq:conf-action-psi} at the second order in $\psi$.

\subsection{Embedding into MAG: hook-symmetric degrees of freedom}\label{sect:MaptoMags-HS}

Now we follow the steps to work out the flat-space kinetic terms for the hook-symmetric trace-free field, which describes one of the two purely tensorial components of the non-metricity. In four spacetime dimensions we have the action
\begin{align}
S_I [g,\psi,\phi] &= \int \sqrt{g}  \left\{
\frac{1}{2} \mathring{\nabla}_\lambda \psi^\rho{}_{\mu\nu} \mathring{\nabla}^\lambda \psi_\rho{}^{\mu\nu}
- \frac{4}{3} \mathring{\nabla}_\mu \psi^\mu{}_{\alpha\beta} \mathring{\nabla}_\nu \psi^{\nu\alpha\beta}
- \frac{2}{3} \mathring{\nabla}_\rho \psi^{\alpha\rho\beta} \mathring{\nabla}_\lambda \psi_\alpha{}^\lambda{}_\beta
+ \frac{1}{2} \mathring{\mathcal{J}} \psi^2 \right. \\\nonumber
& \qquad \left.  + \mathring{K}_{\mu\nu} \psi^\mu{}_{\alpha\beta} \psi^{\nu\alpha\beta} + 2 \mathring{K}_{\mu\nu} \psi_\alpha{}^\mu{}_\beta \psi^{\alpha\nu\beta}  +  \beta_1 \mathring{W}^{\alpha\mu\beta\nu} \psi^\lambda{}_{\alpha\beta} \psi_{\lambda\mu\nu} + \beta_2 \mathring{W}^{\alpha\beta\mu\nu} \psi_\alpha{}^\lambda{}_\beta \psi_{\mu\lambda\nu} \right. \\\nonumber
& \qquad \left.  + \frac{1}{2} \left( \partial_\mu \phi \right)^2 + \frac{1}{12} \phi^2 \mathring{R} - \frac{\lambda}{4!} \phi^4 - \frac{\xi}{2} \phi^2 \psi^2 + \frac{\alpha_3}{2} \mathring{W}^2_{\mu\nu\rho\sigma} \right\} \, ,
\end{align}
which we also coupled to a classically conformally covariant scalar field, $\phi$, and to a manifestly covariant Weyl squared term for generality.
From the action we read off the momentum-space kinetic terms in the flat-space limit using \cite{Baldazzi:2021kaf}
\begin{align}
& a(2^+)_{gg} = - \frac{q^2 \phi_0^2}{32} - q^4 \frac{\alpha_3}{4} \, ,
&& a(2^+)_{\psi\psi} = 2q^2 - 4\xi \phi_0^2 \, , \nonumber
\\
& a(2^-)_{\psi\psi} = - 2 q^2 - 4 \xi \phi_0^2 \, ,
&& a(1^+)_{\psi\psi} = - \frac{2q^2}{3} - 4 \xi \phi_0^2 \, ,
\\
& a(1^-)_{\psi\psi} =\frac{14 q^2}{9} - 4\xi \phi_0^2 \, ,
&& a(0^+)_{gg} = q^4 \frac{4 \alpha_3}{3} + \frac{q^2 \phi_0^2}{24} \, .
\nonumber
\end{align}
We note that the behavior of a given spin-parity sector is completely analogous to the one that we have found in the hook anti-symmetric case. In fact, the mixing in the $2^+$ sector is ruled out in a similar way. Therefore, the hook symmetric and hook anti-symmetric trace-free tensors not only have the same dimension and spin-parity components, but their conformal actions also share the same ghosts and tachyons when expanded around flat-space.

\subsection{Cartan's action for the hook-symmetric traceless tensor}

Now we want to repeat the construction of an interacting MAG action for the traceless hook-symmetric tensor in $d=4$. In this case, we start by considering the connection which is symmetric and noncompatible with the metric \eqref{eq:connection-psi}.
The strategy is similar to the antisymmetric case, but slightly more involved because the curvature tensor has less symmetries. We now have six terms with the curvature tensor squared. Furthermore, it is now possible to define two inequivalent ``Ricci'' tensors
\begin{equation}
 (Ric_1{})_{\mu\nu}=R^\rho{}_{\mu\rho\nu}\,,
\qquad
 (Ric_2{})_{\mu\nu}=R_\mu{}^\rho{}_{\rho\nu} \, ,
\end{equation}
where the first construction is the ``natural'' generalization of Ricci tensor of Riemannian geometry, in that it does not involve the metric to raise and lower indices.
We can write three inequivalent terms quadratic in the Ricci tensors. Among all possible terms, we choose those that return inequivalent contributions once we specialise to our limit, that is, when we impose that nonmetricity is completely determined by $\psi$. 

We thus consider the action
\begin{align}
 S_{\rm MAG}[g,\psi] & = \int \sqrt{g}   \Big\{ p R \, Q^{\alpha\beta\gamma}Q_{\alpha\beta\gamma} + q R^{\mu\nu\rho\sigma}R_{\mu\nu\rho\sigma} + r R^{\mu\nu\rho\sigma}R_{\rho\sigma\mu\nu}  + s R^{\mu\nu\rho\sigma}R_{\mu\rho\nu\sigma}  \nonumber \\
& \qquad + t R^{\mu\nu\rho\sigma}R_{\rho\nu\mu\sigma} + u R^{\mu\nu\rho\sigma}R_{\rho\mu\nu\sigma} +w R^{\mu\nu\rho\sigma}R_{\nu\rho\mu\sigma} + x (Ric_2{})^{\mu\nu}(Ric_2{})_{\mu\nu} \\
& \qquad + y (Ric_2{})^{\mu\nu}(Ric_2{})_{\nu\mu} +z(Ric_1{})^{\mu\nu}(Ric_1{})_{\mu\nu} \Big\} \,. \nonumber 
\end{align}
Rewriting this action in the Einstein form, imposing $Q_{\mu\nu\rho} =\psi_{\mu\nu\rho}$ and keeping only the contributions quadratic in the hook-symmetric traceless tensor, we can completely fix the constants in the MAG action that reproduce our Weyl-invariant action for $\psi$. The explicit solution is
\begin{align} 
 & p=-\frac{107}{168} \, , 
 && r=-\frac{5}{7}  \,,
 && w=-\frac{1}{2}+2q+s+t-u\, , 
 && x=-\frac{11}{21}  \,,  \\
 \nonumber\\
 &  y=-\frac{29}{42} \, , 
 && z=-\frac{13}{28} \,,
 && \beta_1=-\frac{43}{56} \, , 
 && \beta_2=-\frac{67}{28} .
 \nonumber
\end{align}
Similarly to the case presented in the previous section, this also fixes the couplings $\beta_1$ and $\beta_2$ with the Weyl tensor, which had not previously been fixed by simply requiring Weyl invariance. This result tells us that the action that we stared with has a redundancy once we impose $Q_{\mu\nu\rho} =\psi_{\mu\nu\rho}$, because some constants are not determined. We could thus choose $q=s=t=u=0$ in the MAG action and arrive at the conclusion that $w=-\frac{1}{2}$.
As before, we have found a geometrical MAG action in the Cartan form that can be expanded as
\begin{equation}
 S_{\rm MAG}[g,\psi] = S_{\rm conf}[g,\psi] + S_{\rm int}\,,
\end{equation}
where the terms $S_{\rm int}$ are higher order interactions in the tensor $\psi$ with schematic forms $\psi^2\partial \psi$ and $\psi^4$.

\section{Conclusion}

In this paper we have considered the conformal coupling of hook-symmetric and hook-antisymmetric traceless fields, and, in particular, we have found conformally covariant operators on their spaces and actions that are quadratic in the fields. These mixed-symmetry tensors are naturally interpreted as two algebraic irreducible tensorial components of the decompositions of torsion and nonmetricity, which are the distinguishing tensors of a metric-affine theory of gravity. Having found the conformal coupling for these fields, we have then written their actions in forms that highlight their properties as metric-affine models. A future perspective of our wouk would be to push the analysis even further: since the conformal actions for all the algebraic irreducible components of the post-Riemannian fields are known, one may study full conformally invariant actions that are quadratic in all these fields. The interesting question, which we addressed for the specific models at hand but is not answered in general, would be to check which of these models have ghosts or tachyons, as well as to classify their spin-parity sectors.
Some spin-parity sectors are the most troublesome and should be studied; for example, it would be interesting to check if ghosts and tachyons arise in the $2^+$ and $1^-$ sectors also in these more general scenarios.

The relative simplicity of our conformal action has allowed us to avoid
a further complication, which is the possible mixing between fields with the same spin-parity: for example, it would be possible to write down three independent ``kinetic terms'' coupling a hook-symmetric tensor and a hook-antisymmetric tensor.

As we have seen in our analysis, there is no mixing between the $2^+$ components of the metric perturbation and the longitudinal modes of the hook-(anti)symmetric tensors. This is due to the absence of conformally invariant interaction terms of the form $\mathring{Ric} \mathring{\nabla} T$ or $\mathring{Ric} \mathring{\nabla} Q$. Therefore, it is not possible to ``cure'' the pathological behavior of the $2^+$ component of the metric perturbation by a suitable coupling to the torsion and the nonmetricity, at least as far as we are concerned with the flat-space limit. However, we may try and relax the assumption of conformal invariance, say by imposing that it is realized only in the limit of maximally symmetric spacetimes. In this way, we could be able to write down such interaction terms and it could be possible, at least in principle, to look for specific regions in parameter space in which the theory becomes free of ghosts and tachyons. We remark that the specialization to maximally symmetric spacetimes not only simplifies considerably the Lagrangian, but is also well-suited for the application to cosmological models of the very early universe.

We have also shown that the action for conformally coupled traceless hook-antisymmetric tensor cannot be written in a Cartan for which is just quadratic in both the ``full'' curvature tensor (the curvature of the non-Riemannian connection) and torsion \cite{Sezgin:1981xs,Neville:1978bk,Neville:1979rb,Sezgin:1979zf,Sezgin:1980tp}. Notice that this feature only appears when we deal with the metric MAGs. These particular subclasses of metric-affine models are of interest because they are motivated by the analogy with Yang-Mills theories: in this case one views the full curvature tensor as the curvature of the local $GL(4)$ connection, but also the torsion as the curvature of the frame-field and the nonmetricity as the curvature of a local metric tensor $g_{ab}$. We emphasize that this analogy is of a rather heuristic origin. Indeed, curvature tensors are usually written in terms of one affine-connection only, and they do not depend on another field. Moreover, both torsion and nonmetricity have the ``wrong'' parity when it comes to interpret them as field strengths, whereas the full curvature tensor still does have the right parity. Therefore, on these grounds, we are tempted to claim that these models that are analog to Yang-Mills might be an oversimplification, and that we should rather think of torsion and nonmetricity as (a particular combination of) the gauge-covariant parts of the full affine-connection of spacetime. Some simplifications may still be required, in the form of some Curtright-like invariances of the full action functional, which we have also touched in the paper. In particular, we may try and remove troublesome longitudinal modes which give rise to ghosts and tachyons using generalized gauge symmetries in order to produce models of more interesting phenomenological value.

\smallskip

\paragraph*{Acknowledgments.} We are very grateful to
M.~Grigoriev, M.~Ponds, J.~Queva, I.~Shapiro and M.~A.~Vasiliev
for important comments and clarifications.

\appendix

\section{Algebraic reduction of the independent terms}\label{sect:bases}

In this appendix we show how many independent terms are needed when constructing the ansatze for the hook-antisymmetric and hook-symmetric actions of Sects.~\ref{sect:hook-antisymmetric} and \ref{sect:hook-symmetric}.
We concentrate on the hook-antisymmetric case for this presentation, but the hook-symmetric case works analogously except for minor differences that we point out at the end.

The first step is the construction of all the independent expressions of two $\kappa$ tensors, in which two indices are not contracted. Since $\kappa$ is traceless, each of the two $\kappa$ tensors will have an uncontracted index. There are, in principle, four possible contractions
\begin{equation}
 {M_1}^{\mu\lambda} = \kappa^\mu{}_{\nu\rho} \kappa^{\lambda\nu\rho} \, , \qquad {M_2}^{\mu\lambda} = \kappa^{\mu\nu\rho} \kappa_\nu{}^\lambda{}_\rho \, , \qquad {M_3}^{\mu\lambda} = \kappa^{\nu\mu\rho} \kappa_\nu{}^\lambda{}_\rho \, , \qquad {M_4}^{\mu\lambda} = \kappa^{\nu\mu\rho} \kappa_\rho{}^\lambda{}_\nu \, .
\end{equation} 
First, it is easy to see that the second contraction is actually a multiple of the first one, using first the antisymmetry of the contracted indices and then the cycle property of hook-antisymmetry
\begin{align}\label{eq:kappa-contr-1}
 \qquad {M_2}^{\mu\lambda} & = \kappa^{\mu\nu\rho} \kappa_\nu{}^\lambda{}_\rho = \frac{1}{2} \kappa^{\mu\nu\rho} \left( \kappa_\nu{}^\lambda{}_\rho - \kappa_\rho{}^\lambda{}_\nu \right) = \frac{1}{2} \kappa^{\mu\nu\rho} \left( \kappa_\nu{}^\lambda{}_\rho + \kappa^\lambda{}_{\nu\rho} + \kappa_{\nu\rho}{}^\lambda \right) =\frac{1}{2} \kappa^{\mu\nu\rho} \kappa^\lambda{}_{\nu\rho} = \frac{1}{2} {M_1}^{\mu\lambda}\, .
\end{align}
Then, we show that the fourth contraction is a linear combination of the first and the third one by cycling the indices in one of the two $\kappa$ tensors and using Eq.~\eqref{eq:kappa-contr-1}
\begin{align}\label{eq:kappa-contr-2}
 {M_4}^{\mu\lambda} & = \kappa^{\nu\mu\rho} \kappa_\rho{}^\lambda{}_\nu = - \kappa^{\nu\mu\rho} \left( \kappa^\lambda{}_{\nu\rho} + \kappa_{\nu\rho}^\lambda \right) = - \frac{1}{2} \kappa^{\mu\nu\rho} \kappa^\lambda{}_{\nu\rho} + \kappa^{\nu\mu\rho} \kappa_\nu{}^\lambda{}_\rho = - \frac{1}{2} {M_1}^{\mu\lambda} + {M_3}^{\mu\lambda} \, .
\end{align}
We cannot further reduce the basis, therefore there are only two ways to contract two $\kappa$ tensors with a trace-free symmetric tensor. We also notice that the terms share the same trace,
\begin{equation}
 {M_1}^\mu{}_\mu = {M_3}^\mu{}_\mu \, ,
\end{equation}
therefore there is only one independent complete contraction of two $\kappa$ tensors.
The exact same logic can be extended straightforwardly to the hook-symmetric case, the only differences being signs and relative factors which parametrize the linear dependences (see below).
This analysis explains the completeness and nonredundancy of our starting actions $S_0$ and $S_1$ in Sects.~\ref{sect:hook-antisymmetric} and \ref{sect:hook-symmetric}.

Now we concentrate on contractions of two $\kappa$ tensors which are rank-$4$ tensors, needed to construct the independent couplings with $\mathring{W}_{\mu\nu\rho\theta}$. Using the similar algebraic manipulations, it is not difficult to show that there are only three inequivalent contractions
\begin{equation}
{K_1}_{\mu\nu\rho\sigma} = \kappa^\lambda{}_{\mu\nu} \kappa_{\lambda\rho\sigma} \, , \, \qquad {K_2}_{\mu\nu\rho\sigma} = \kappa_{(\mu}{}^\lambda{}_{\nu)} \kappa_{(\rho|\lambda|\sigma)} \, ,  \qquad {K_3}_{\mu\nu\rho\sigma} = \kappa^\lambda{}_{\mu\nu} \kappa_{(\rho|\lambda|\sigma)} \, .
\end{equation}
The first two contractions are antisymmetric and symmetric in the two pairs of indices, respectively, and the pairs can also be exchanged. The third contraction has instead less symmetries. Naively, there are two ways to contract each of the rank-$4$ contractions with the Weyl tensor, giving a total of six possible terms
\begin{align}\label{eq:kappa-Weyl1}\nonumber
 & {K_1}_{\mu\nu\rho\sigma} \mathring{W}^{\mu\nu\rho\sigma} \, ,
 &&  {K_1}_{\mu\nu\rho\sigma} \mathring{W}^{\mu\rho\nu\sigma} \, ; \\
 & {K_2}_{\mu\nu\rho\sigma} \mathring{W}^{\mu\rho\nu\sigma} \, , 
 && {K_2}_{\mu\nu\rho\sigma} \mathring{W}^{\mu\nu\rho\sigma} \, ; \\\nonumber
 & {K_3}_{\mu\nu\rho\sigma} \mathring{W}^{\mu\nu\rho\sigma} \, ,
 && {K_3}_{\mu\nu\rho\sigma} \mathring{W}^{\mu\rho\nu\sigma} \, .
\end{align}
First, we notice that the first contraction of $K_3$ with the Weyl tensor identically vanish because of the symmetries of the two tensors. The same is true also for the second contraction involving the $K_2$ rank-$4$ tensor. Then, the second contraction of $K_1$ can be rewritten in terms of the first one
\begin{align}
 {K_1}_{\mu\nu\rho\sigma} \mathring{W}^{\mu\rho\nu\sigma} & = \frac{1}{2} {K_1}_{\mu\nu\rho\sigma} \left( \mathring{W}^{\mu\rho\nu\sigma} + \mathring{W}^{\nu\mu\sigma\rho} + \mathring{W}^{\nu\sigma\rho\mu} \right) = \frac{1}{2} {K_1}_{\mu\nu\rho\sigma} \mathring{W}^{\mu\nu\rho\sigma} \, .
\end{align}
Finally, the second contraction of $K_3$ is proportional to the first by virtue of symmetries and Bianchi identities
\begin{align}
 {K_3}_{\mu\nu\rho\sigma} \mathring{W}^{\mu\rho\nu\sigma} & = \frac{1}{2} {K_3}_{\mu\nu\rho\sigma} \left( \mathring{W}^{\mu\rho\nu\sigma} - \mathring{W}^{\nu\rho\mu\sigma} \right) =  \frac{1}{2} {K_3}_{\mu\nu\rho\sigma} \left( \mathring{W}^{\mu\rho\nu\sigma} + \mathring{W}^{\nu\mu\sigma\rho} + \mathring{W}^{\nu\sigma\rho\mu} \right) = \frac{1}{2} {K_3}_{\mu\nu\rho\sigma} \mathring{W}^{\mu\nu\rho\sigma} = 0 \, .
\end{align}
Therefore, we have chosen a basis for the conformal action \eqref{eq:Conf-Action-Kappa} that includes two independent contractions with the Weyl tensor
and that can be easily shown to be linear combinations of the above.

To highlight the differences that arise in this construction for the hook-symmetric terms, we start building the rank-$2$ tensors with two $\psi$
\begin{equation}
 {N_1}^{\mu\lambda} = \psi^\mu{}_{\nu\rho} \psi^{\lambda\nu\rho} \, , \qquad {N_2}^{\mu\lambda} = \psi^{\mu\nu\rho} \psi_\nu{}^\lambda{}_\rho \, , \qquad {N_3}^{\mu\lambda} = \psi^{\nu\mu\rho} \psi_\nu{}^\lambda{}_\rho \, , \qquad {N_4}^{\mu\lambda} = \psi^{\nu\mu\rho} \psi_\rho{}^\lambda{}_\nu \, .
\end{equation}  
Proceeding as before it is immediate to see that
\begin{equation}
 {N_2}^{\mu\lambda} = -\frac{1}{2}{N_1}^{\mu\lambda}  \, , \qquad {N_4}^{\mu\lambda} = \frac{1}{2}{N_1}^{\mu\lambda}+{N_3}^{\mu\lambda},
\end{equation} 
and that we have just one scalar combination (that can be checked also by a direct computation).
Then we consider independent rank-$4$ tensors with two $\psi$. Again, there are three
\begin{equation}
{\Psi_1}_{\mu\nu\rho\sigma} = \psi^\lambda{}_{\mu\nu} \psi_{\lambda\rho\sigma} \, , \, \qquad {\Psi_2}_{\mu\nu\rho\sigma} = \psi_{[\mu}{}^\lambda{}_{\nu]} \psi_{[\rho|\lambda|\sigma]} \, ,  \qquad {\Psi_3}_{\mu\nu\rho\sigma} = \psi^\lambda{}_{\mu\nu} \psi_{[\rho|\lambda|\sigma]} \, ,
\end{equation}
so, once more, there are six possible contractions with the Weyl tensor.
However, we have that ${\Psi_1}_{\mu\nu\rho\sigma} \mathring{W}^{\mu\nu\rho\sigma}=
{\Psi_1}_{\mu\nu\rho\sigma} \mathring{W}^{\mu\rho\nu\sigma}=0$, and also
\begin{align}\label{eq:psi-Weyl1}\nonumber
 {\Psi_2}_{\mu\nu\rho\sigma} \mathring{W}^{\mu\nu\rho\sigma}   = 2{\Psi_2}_{\mu\nu\rho\sigma} \mathring{W}^{\mu\rho\nu\sigma}
 \,,\qquad
 {\Psi_3}_{\mu\nu\rho\sigma} \mathring{W}^{\mu\rho\nu\sigma}   = \frac{1}{2}{\Psi_3}_{\mu\nu\rho\sigma} \mathring{W}^{\mu\nu\rho\sigma}=0\,,
\end{align}
so there are only two independent contractions with the Weyl tensor
as seen in \eqref{eq:conf-action-psi}.

\section{On the relation with gauged Weyl symmetry} \label{sect:relation}

The vectors in the decompositions of torsion and nonmetricity in Sect.~\ref{sect:introduction} are known to be related to (gauged) Weyl geometries in some limits.
We set the stage by noticing that Weyl symmetry can be gauged using an Abelian potential $S_\mu$, so that the covariant derivative
\begin{equation}\label{eq:gauged-weyl-covd}
 \begin{split}
  \hat{\nabla}_\mu v^\nu= \mathring{\nabla}_{\mu} v^\nu + (\delta^\nu_{\rho}S_\mu + \delta^\nu_{\mu}S_\rho- g_{\rho\mu} S^\nu) v^\rho + w_v S_\mu v^\rho\,,
 \end{split}
\end{equation}
transforms covariantly under the gauged Weyl transformations
\begin{equation}
 \begin{split}
  g_{\mu\nu} \to g'_{\mu\nu}={\rm e}^{2 \sigma} g_{\mu\nu}\,, \qquad
  S_\mu \to S'_\mu = S_\mu -\partial_\mu \sigma\,, \qquad
  v^\mu \to v'{}^\mu = {\rm e}^{w_v \sigma} v^\mu\,,
 \end{split}
\end{equation}
where $S_\mu$ has an affine transformation, as discussed, for example, in Ref.~\cite{Iorio:1996ad}. The coupling $w_v$ is the Weyl weight of $v^\mu$ and is the charge of the gauged Weyl transformation. The covariant derivative $\hat{\nabla}$ is not the direct sum of the Levi-Civita and a gauge one because dilatations do not commute with the local Lorentz group, still the semidirect product of the generated group is a subgroup of $GL(d)$ \cite{Sauro:2022hoh}.

The second term of \eqref{eq:gauged-weyl-covd} can be seen as a disformation, which results from a special type of nonmetricity $Q_{\mu\nu\rho}=-2 S_\mu g_{\nu\rho}$, while the third term is a gauge contribution that balances the former when acting on the metric $\hat{\nabla}_\mu g_{\nu\rho}=0$
given that the weight of the metric is $w_{g_{\mu\nu}}=2$.

In order to obtain the correct nonmetricity and disformation from our decompositions \eqref{eq:T-Q-decomposition}, it is sufficient to impose
relations among the nonmetricity vectors and $S_\mu$
\begin{equation}
 \begin{split}
  V_\mu = \frac{1}{d} Z_\mu\,, \qquad S_\mu = -\frac{1}{2} V_\mu\,,
 \end{split}
\end{equation}
telling us that $V_\mu$ functions as $S_\mu$, provided that the vectors $V_\mu$ and $Z_\mu$ are related as above. Notice that, as outlined in the example of conformal action of a vector given in Sect.~\ref{sect:conformal-past}, in $d=4$
the Abelian gauge and conformal symmetries are compatible.

A similar construction can be repeated using the torsion vector $\tau_\mu$ as gauge Weyl potential \cite{Karananas:2015eha,Karananas:2021gco}. The relation is $S_\mu=\frac{1}{d-1}\tau_\mu$, so the torsion vector must transform as $\tau_\mu \to \tau_\mu -(d-1) \partial_\mu \sigma$ under Weyl transformations (we have a different sign in the transformation as compared with \cite{Karananas:2015eha} because the torsion vector is defined differently).
There are two ways to go about this construction. The first is to follow Ref.~\cite{Karananas:2015eha} and use both the coset approach and the inverse Higgs mechanism to solve algebrically a vector component of the covariant differential of the coframe in terms of the fields, giving the above relation.
A second approach involves the use of a special projective symmetry, which leaves invariant null autoparallels, as done in Ref.~\cite{Sauro:2022hoh}.

Gauged Weyl invariance was originally proposed by Weyl himself as an attempt to unify electromagnetism and gravity, by extending the covariance of General Relativity to a larger covariance under local dilatations, while the gauge group remains a subgroup of $GL(d)$. The attempt was ultimately a failure, but has lead to the first gauge theory ever constructed \cite{Weyl:1918ib}.
In practice, Weyl symmetry can always be made covariant through the use of a Weyl vector potential. As such, it is not particularly difficult to find Weyl invariant actions by just replacing $\mathring{\nabla} \to \hat{\nabla}$,
and the resulting actions are not very constrained \cite{Iorio:1996ad},
but are also receiving renewed attention in the past years \cite{Ghilencea:2021lpa,Ghilencea:2018thl}.
However, one important point to make is the one of Ref.~\cite{Sauro:2022chz},
where it was noted that the connection $\hat{\nabla}$ can be complemented by other Weyl invariant tensors including, in particular, $H^\mu{}_{\nu\rho}$ and $\kappa^\mu{}_{\nu\rho}$ of \eqref{eq:T-Q-decomposition}.


\end{document}